\documentclass[preprint,12pt]{elsarticle}

\usepackage{array}
\usepackage{soul}
\usepackage{amsthm}
\newtheorem{assumption}{Assumption}
\newtheorem{theorem}{Theorem}
\usepackage{multirow}
\usepackage{setspace}
\usepackage{moreverb}
\usepackage{amsmath}
\usepackage{amsfonts}
\usepackage{amssymb}
\usepackage{graphicx}
\usepackage{caption}
\usepackage{subcaption}
\usepackage{enumerate}
\usepackage{xcolor}
\usepackage{natbib} 
\usepackage{url} 
\usepackage{bm}
\usepackage{hyperref}
\usepackage{cleveref}
\usepackage{multicol}

\usepackage{lipsum}

\hypersetup{breaklinks=true,
            pdfauthor={},
            pdfkeywords = {},
            pdftitle={},
            colorlinks=true,
            citecolor=blue,
            urlcolor=blue,
            linkcolor=magenta,
            pdfborder={0 0 0}}

\usepackage{algorithm,algpseudocode}

\newcounter{algsubstate}
\renewcommand{\thealgsubstate}{\alph{algsubstate}}

\newcommand\BibTeX{{\rmfamily B\kern-.05em \textsc{i\kern-.025em b}\kern-.08em
\kern-.1667em\lower.7ex\hbox{E}\kern-.125emX}}

\begin{document}

\begin{titlepage}
    \centering
    \vspace*{1cm}

    \Huge
    \textbf{Enhancing Approximate Modular Bayesian Inference by Emulating the Conditional Posterior}

    
    \vspace{1.5cm}
    \Large
    Grant Hutchings\textsuperscript{a,b}\footnote{Corresponding author: Email: grant.hutchings@lanl.gov, Present address: P.O. Box 1663 Los Alamos, NM 87545.},
    Kellin N. Rumsey\textsuperscript{a},
    Derek Bingham\textsuperscript{b},
    Gabriel Huerta\textsuperscript{c,d}

    \vfill

    \textsuperscript{a}Statistical Sciences, Los Alamos National Laboratory, NM, United States\\
    \textsuperscript{b}Department of Statistics, Simon Fraser University, BC, Canada\\
    \textsuperscript{c}Statistical Sciences, Sandia National Laboratories, NM, United States\\
    \textsuperscript{d}Department of Mathematics and Statistics, University of New Mexico, NM, United States

    \vspace{1.5cm}

    \large
    \today

\end{titlepage}



\title{Enhancing Approximate Modular Bayesian Inference by Emulating the Conditional Posterior}

\author[cor1,1,2]{Grant Hutchings}

\author[1]{Kellin N. Rumsey}

\author[2]{Derek Bingham}

\author[3,4]{Gabriel Huerta}


\address[1]{Statistical Sciences, Los Alamos National Laboratory, NM, United States}
\address[2]{Department of Statistics, Simon Fraser University, BC, Canada}
\address[3]{Statistical Sciences, Sandia National Laboratory, NM, United States}
\address[3]{Department of Mathematics and Statistics, University of New Mexico, NM, United States}

\cortext[cor1]{Corresponding author: Email: grant.hutchings@lanl.gov (Grant Hutchings). Present address: P.O. Box 1663 Los Alamos, NM 87545.}

\begin{abstract}
In modular Bayesian analyses, complex models are composed of distinct modules, each representing different aspects of the data or prior information. In this context, fully Bayesian approaches can sometimes lead to undesirable feedback between modules, compromising the integrity of the inference. This paper focuses on the ``cut-distribution" which prevents unwanted influence between modules by ``cutting" feedback. The multiple imputation (DS) algorithm is standard practice for approximating the cut-distribution, but it can be computationally intensive, especially when the number of imputations required is large. An enhanced method is proposed, the Emulating the Conditional Posterior (ECP) algorithm, which leverages emulation to increase the number of imputations. Through numerical experiment it is demonstrated that the ECP algorithm outperforms the traditional DS approach in terms of accuracy and computational efficiency, particularly when resources are constrained. It is also shown how the DS algorithm can be improved using ideas from design of experiments. This work also provides practical recommendations on algorithm choice based on the computational demands of sampling from the prior and cut-distributions.
\end{abstract}

\begin{highlights}
\item A new algorithm is proposed for approximating the cut-distribution which enhances multiple imputation when computational resources are limited.
\item The use of experimental design to reduce variability and improve accuracy of the multiple imputation approach is discussed.
\end{highlights}


\maketitle

\section{Introduction}

Modular statistical inference refers to the problem of conducting inference when a statistical model is composed of distinct components, or modules, which must be combined to perform an analysis \citep{jacob2017BetterTogether, liu2009, plummer2015}. These modules are often heterogeneous, possibly coming from different communities or data sources, and the level of {\it a priori} trust may vary across different components of the model. In this case, it is often best to forego a fully Bayesian analysis, in favor of analyses that treat modules (at least partially) separately. This idea is explored, for example, in the context of computer models in the foundational work \cite{liu2009}, where the authors give several reasons for considering a modular Bayesian analysis, including
\begin{itemize}
\setlength\itemsep{-0.5em}
\item to avoid contamination of a trustworthy module by a suspect module
\item for better scientific understanding and future scientific development of the modules
\item due to lack of identifiability between parameters across modules
\item numerical issues (e.g. poor MCMC mixing)
\end{itemize}

Consider a statistical model with data sources $\bm y$ and $\bm z$, where the probability density function for $\bm y$ depends on parameters $\bm \alpha \in A \subseteq \mathbb R^p$, and the probability density function for auxiliary data $\bm z$ depends on parameters $\bm\gamma \in \Gamma \in\mathbb R^q$. The auxiliary information $\bm z$ can be explicit in the form of data, but can also be implicit, such as expert opinion which is used to construct prior distributions. Since $\bm z$ does not depend on $\bm\alpha$, the full marginal posterior distribution for $\bm \alpha$ can be written as 
\begin{equation}
\label{eq:post}
\pi(\bm\alpha|\bm\ y) = \int_{\Gamma}\pi(\bm\alpha, \bm\gamma|\bm y, \bm z) d\bm\gamma = \int_{\Gamma} \pi(\bm\alpha|\bm\gamma, \bm y)  \pi(\bm\gamma|\bm y, \bm z) d\bm\gamma.
\end{equation}
The far right hand side of \cref{eq:post}, while not very useful in practice, illustrates that the marginal posterior of $\bm \alpha$ can be viewed as taking the conditional posterior density of $\pi(\bm\alpha|\bm\gamma,\bm y)$ and averaging it across the marginal posterior density $\pi(\bm\gamma|\bm y, \bm z)$. For any of the reasons described above, it is often the case that the influence of $\bm y$ can have an undesirable influence on the estimate of the posterior distribution for $\bm\gamma$, and more trustworthy inference can be obtained by ``cutting" the feedback from $\bm y$ to $\bm\gamma$ so that inference for $\bm \gamma$ is conditional only on $\bm z$.

One common version of a Bayesian cut procedure is the so-called \textit{plug-in} approach, in which feedback from $\bm y$ to $\bm \gamma$ is cut by treating $\bm\gamma$ as fixed and known. Denoting the Dirac delta function by $\delta(\cdot)$, the ``plug-in" distribution is given by
\begin{equation}
\label{eq:plug}
\begin{aligned}
\pi_{\text{plug}}(\bm\alpha|\bm\ y) &= \int_{\Gamma} \pi(\bm\alpha|\bm\gamma, \bm y) \delta(\bm\gamma - \hat{\bm\gamma}) d\bm\gamma \\[1.2ex]
&= \pi(\bm\alpha|\hat{\bm\gamma}, \bm y), 
\end{aligned}
\end{equation}
where $\hat{\bm\gamma}$ is an estimate of $\bm\gamma$, such as $\hat{\bm\gamma} = \int_{\Gamma} \bm\gamma \pi(\bm\gamma|\bm z) d\bm\gamma$. This idea has been used to varying degrees in many applications \cite{beven2001, arendt2012, brown2018, rumsey2020dealing}. The plug-in approach has several obvious drawbacks however, including potential sensitivity of the posterior distribution of $\bm \alpha$ to the estimated value of $\bm \gamma$ and underestimation of uncertainty in the same distribution. 

It is desirable to have a more rigorous alternative to cutting the influence of $\bm y$ on $\bm\gamma$ directly, while still maintaining the forward propagation of uncertainty. The resulting distribution, known as the \textit{cut-distribution} or \textit{cut-posterior}, has been studied in recent years, including \cite{plummer2015, jean2009, rumsey2020methods, jacob2017BetterTogether, jacob2020UnbiasedMarkov} and is given by 
\begin{equation}
\label{eq:cutdistr}
\pi_{\text{cut}}(\bm\alpha|\bm\ y) = \int_{\Gamma} \pi(\bm\alpha|\bm\gamma, \bm y) \bm \pi(\bm\gamma| \bm z) d\bm\gamma.
\end{equation}
Notice that $\pi(\bm \gamma|\bm y, \bm z)$ from \cref{eq:post} is replaced by $\pi(\bm \gamma | \bm z)$ in \cref{eq:cutdistr}. 

In some applications, the dependence on $\bm z$ is conceptual (i.e., prior information) and can be dropped to simplify the notation. In that scenario, the cut-distribution can be viewed as averaging the conditional posterior distribution of $\bm\alpha$ (given $\bm\gamma$) across the prior distribution of $\bm\gamma$. More generally, it is usually difficult to compute \cref{eq:cutdistr} in closed form, and we must settle for an approximation. 
The most straightforward approach, which we refer to as direct sampling (DS; sometimes called multiple imputation or parallel cutting), can be time-intensive when MCMC is needed to sample from the conditional posterior distribution, especially if a large number of imputations are required for accurate approximation.


The primary contribution of this paper is proposing a new approach for enhancing the DS algorithm, especially when only a small number of imputations are possible. Each imputation targets the conditional posterior of $\bm \alpha$ (given $\bm \gamma$) for a specific $\bm \gamma$. The proposed approach fits a parametric approximation to each of these conditional distributions, and uses a statistical emulator to predict the parametric form at unsampled values of $\bm \gamma$. The emulator is used to effectively increase the number of imputations with minimal computational effort. This approach can lead to drastically better approximations of the cut-distribution when the number of imputations is limited \citep{rumsey2020methods}. In a different context, similar ideas have recently been introduced \cite{GPalphagamma2023, vernon2023bayesian}.


A secondary contribution of this work is an exploration of the traditional DS algorithm in practice. Specifically, the use of ideas from design of experiments (DOE) to improve approximations of the cut-distribution with DS is proposed.


The remainder of this work is organized as follows. In \cref{sec:cut_review}, a summary of current methodologies for approximating the cut-distribution is given. In \cref{sec:ECP}, the new approach which rapidly converges to the cut-distribution in many useful cases is proposed. This procedure is referred to as the ECP (emulating the conditional posterior) algorithm and is compared to DS using a simple test problem. In \cref{sec:DS_DOE} it is shown how the DS algorithm can be improved using DOE, but that it is sill not competitive with ECP on the test problem for small computational budgets. In \cref{sec:examples}, a more complex numerical example is presented which shows that the ECP algorithm makes better use of its computational budget than DS. In \cref{sec:practical} the computational complexity of the proposed approach discussed, and recommendations are given for sampling from the cut-distribution in a number of scenarios. Concluding remarks are given in \cref{sec:discussion}.

\section{Approximating the Cut-Distribution: A Review}\label{sec:cut_review}

\subsection{Direct Sampling}

Close examination of the cut-distribution of \cref{eq:cutdistr} suggests that it can be viewed as the conditional posterior distribution $\pi(\bm\alpha|\bm\gamma,\bm y)$, averaged over $\pi(\bm\gamma|\bm z)$. Although $\pi(\bm\gamma|\bm z)$ is technically a posterior distribution, it can be considered a prior distribution with respect to the module with data $\bm y$, and thus we will often refer to it as such. Viewing the cut-distribution in this way suggests a straightforward two-step algorithm that works by sampling $\bm\gamma$ from its prior distribution and then, treating this value as fixed and known, drawing $m$ samples of $\bm \alpha$ from its conditional posterior distribution $\pi(\bm\alpha|\bm\gamma, \bm y)$. This process is repeated $L$ times and the combined $M=mL$ samples can be treated as $M$ draws from the cut-distribution. We refer to this procedure as direct sampling (DS) and it is considered to be the gold standard approach for sampling from the cut-distribution \cite{plummer2015,jacob2017BetterTogether}.

If MCMC is used for inference on $\bm \alpha$, the samples generated by DS can be considered exact (asymptotically in the length of the MCMC chains) when $m=1$, but this approach may not efficiently use computational resources, as it provides only $M=L$ draws from the cut-distribution. In practice, generating $m>1$ draws from $\pi(\bm\alpha|\bm\gamma, \bm y)$ often requires little additional computation, especially when the MCMC burn-in phase is significant. However, if the ratio $m/L$ becomes too large, the accuracy of the approximation to the cut-distribution can be negatively effected due to sample dependence (see, for example \cref{fig:plummer_post}).

The number of MCMC runs ($L$) is often constrained by computational resources, and while DS can yield a good approximation for large enough $L$, it can become prohibitively expensive in more challenging problems, particularly when parallel processing is limited.

\subsection{Other Methods}

The naive cut (NC) algorithm refers to a modified MCMC procedure where at each iteration we sample (i) $\bm\alpha|\bm y,\bm\gamma$ and (ii) $\bm\gamma|\bm z$, bypassing the usual Bayesian update for the samples of $\bm\gamma$. This approach is computationally convenient and was used in various applications \cite{liu2009, zigler2013model, brown2018}, but \cite{plummer2015} demonstrated that it often fails to converge to the correct distribution. While modifications to the NC algorithm can address this, they generally require full specification of the marginal likelihood of $\bm y|\bm\gamma$, which is seldom available analytically.

The tempered cut (TC) algorithm, introduced by \cite{plummer2015}, addresses the shortcomings of the NC approach. The TC algorithm adjusts the MCMC sampler to approximate the cut-distribution by exploiting limiting cases where NC converges correctly. Each iteration of the sampler involves sampling $\bm\gamma_t |\bm z$ and then sampling $\bm\alpha$ along a sequence of $K$ intermediate values of $\bm\gamma$ from $\bm\gamma_{t-1}$ to $\bm\gamma_t$. Only the final sample of $\bm\alpha$ is retained. For sufficiently large $K$, TC can approximate the cut-distribution well, but determining an appropriate $K$ is empirical and problem-specific, often leading to a significant increase in computational cost compared to NC. 


A modified TC algorithm is developed in \cite{liu2022stochastic}, which improves on the original TC by enabling parallelization of the internal Markov chain. Although this can provide some speed advantages, the algorithm can still become computationally prohibitive, as the number of likelihood evaluations increases with each iteration, often surpassing the available parallel computing resources.

In \cite{jacob2020UnbiasedMarkov}, the authors demonstrate how coupled MCMC chains can be used to construct unbiased estimators for the expected value of a function with respect to an unnormalized target density in finite time, providing a direct way to estimate the cut-distribution. Unlike methods that rely on approximations, this approach directly targets the cut-distribution using MCMC. A key drawback is that the meeting time of the coupled chains can be prohibitively long, making the method impractical under fixed computational budgets. Furthermore, the method estimates the cut-distribution at various values of $\bm\alpha$ with uncertainty. It is unclear how select the best $\bm\alpha$ values and how to properly combine these estimates with their uncertainties into a proper density estimate.


In \cite{yu2023variational}, the authors show that variational inference can be used to efficiently approximate the cut-distribution, offering a much faster alternative to MCMC-based methods. A mean-field variational approximation is used, but is shown to underestimate uncertainty, and the results are unsatisfactory when strong dependence exists between variables. A fixed form variational approximation is developed to overcome this, but still underestimates uncertainty, and may fail to capture the posterior mode.




In the next section we propose a method which overcomes some of the challenges discussed above.

\section{Emulating the Conditional Posterior}\label{sec:ECP}

In this section, we describe an approach for enhancing the DS procedure, leading to a better approximation given similar computational budgets. For illustration, we consider the following problem described generally in \cite{liu2009}. The problem is one of multiple data sources, where a small amount of unbiased data and a large amount of biased data are available to inform a posterior distribution for a parameter of interest. The problem is adapted and presented in a new form to facilitate discussion.

\subsection{Diamond in a Box}\label{sec:diamond}

The Diamond in a Box (DB) problem is defined as follows. 
\begin{quote} \label{prob:DB}
 {\it A valuable diamond weighing $\alpha$ grams is contained in a display case weighing $\gamma$ grams. The curator has a scale which, when measuring an object weighing $x$ grams, produces an output of $x + \epsilon_i$ where $\epsilon_i \stackrel{\text{iid}}{\sim} N(0, \sigma^2)$. The curator weights the diamond $n_1$ times before placing it back in its box. The diamond is very valuable, so it must remain its display case, but the curator can weigh the display case with the diamond inside an additional $n_2$ times if needed. The goal is to assess the weight of the diamond.}
\end{quote}

Without any additional information, one can immediately sense that weighing the box and diamond together is a waste of time. Using the information gained from the first $n_1$ trials we can obtain a reasonable estimate of the diamonds weight. While the second set of $n_2$ trials would allow us to estimate the weight of the display case, it provides no additional information regarding the weight of the diamond. Fortunately enough, the curator is also a Bayesian, and they realize that the situation can be improved with the use of prior information. By extensively measuring a collection of similar display cases, they are able to build an informative prior distribution for $\gamma$. This external source of information can be regarded formally as $\bm z$. The idea of modularization here is that $\bm z$ is used to inform the weight of the box, while $\bm y$ is used to inform the weight of the diamond, and the modules are kept separate to avoid contamination by the suspect module. 

This problem can be stated mathematically as
\begin{equation}
\label{eq:DB}
\begin{aligned}
y_i &= 
\begin{cases}
\alpha + \epsilon_i, & i = 1, \ldots, n_1, \\
\alpha + \gamma + \epsilon_i, & i = n_1+1, \ldots, n,
\end{cases} \\[1.2ex]
\epsilon_i &\overset{\text{iid}}{\sim} N(0, \sigma^2), \quad\quad 
\alpha \sim N(\mu_\alpha, \sigma^2_\alpha), \quad\quad
\gamma \mid \bm{z} \sim N(\mu_\gamma, \sigma^2_\gamma),
\end{aligned}
\end{equation}

\noindent where $n = n_1 + n_2$ and typically $n_1 \ll n_2$. Defining $\bar y_1 = \left(\sum_{i=1}^{n_1} y_i\right)/n_1$ and $\bar y_2 = \left(\sum_{i=1}^{n_2} y_{n+1-i}\right)/n_2$, the marginal posterior distribution for $\alpha$ is
\begin{equation}
\label{eq:Bayes_d}
\begin{aligned}
\pi(\alpha \mid \bm y) &= N\Bigg(\alpha \ \Bigg|\ 
\frac{n_1(n_2\sigma^2_\gamma+\sigma^2)(\bar y_1-\mu_\alpha) + n_2\sigma^2(\bar y_2-\mu_\alpha-\mu_\gamma)}
{(n_1+\sigma^2/\sigma_\alpha^2)(n_2\sigma^2_\gamma+\sigma^2) + n_2\sigma^2}, \\
&\hspace{2cm}
\frac{\sigma^2(n_2\sigma_\gamma^2 + \sigma^2)}
{(n_1 + \sigma^2/\sigma_\alpha^2)(n_2\sigma_\gamma^2 + \sigma^2) + n_2\sigma^2}
\Bigg).
\end{aligned}
\end{equation}
Using \cref{eq:cutdistr} directly, the cut-distribution for $\alpha$ can be written as
\begin{equation}
\label{eq:cut_DB}
\begin{aligned}
\pi_\text{cut}(\alpha \mid \bm y) &= N\Bigg(\alpha \ \Bigg|\ 
\frac{\sigma_\alpha^2\left(n_1\bar{y}_1 + n_2(\bar{y}_2 - \mu_\gamma)\right) + \sigma^2\mu_\alpha}
{\sigma_\alpha^2\left(n_1+n_2\right) + \sigma^2}, \\
&\hspace{2cm}
\frac{1}{n_1 + n_2 + \sigma^2/\sigma^2_\alpha}
\left(\sigma^2 + \frac{n_2^2\sigma_\alpha^2\sigma_\gamma^2}
{\sigma_\alpha^2(n_1+n_2) + \sigma^2} \right)
\Bigg)
\end{aligned}
\end{equation}

\subsection{Speeding Up the Direct Sampling Algorithm}\label{sec:speedingup}

We build intuition for the ECP algorithm by examining the simple DB problem. The DS approach, discussed in \cref{sec:cut_review}, can be viewed as a two-stage procedure: (i) sample $\bm\gamma|\bm z$ and (ii) sample $\bm\alpha|\bm\gamma, \bm y$. While the first stage is often trivial, the second stage will usually be more intensive, possibly requiring an expensive MCMC procedure for each iteration. If the conditional posterior distribution of $\bm\alpha$ was known for all values of $\bm\gamma$, and if it was easy to sample, then this two stage sampling procedure would be very fast.

With this idea in mind, we examine the conditional posterior distribution $\pi(\alpha|\gamma,\bm y)$ for the Diamond in a Box problem of \cref{eq:DB}.
\begin{equation}
\label{eq:condpost_DB}
\begin{aligned}
\pi(\alpha \mid \gamma, \bm{y}) &= N\left(B + C\gamma, A\right) \\[1.5ex]
A &= \left(\frac{n}{\sigma^2} + \frac{1}{\sigma_\alpha^2}\right)^{-1}, \\
B &= A\left(\frac{\sum_{i=1}^n y_i}{\sigma^2} + \frac{\mu_\alpha}{\sigma_\alpha^2}\right), \\
C &= \frac{-An_2}{\sigma^2}.
\end{aligned}
\end{equation}

\Cref{eq:condpost_DB} reveals that the conditional posterior distribution of $\alpha$ depends on $\gamma$ in a very simple way. In fact, only a budget of $L=2$ imputations is required to estimate the constants $A$, $B$, and $C$. To do this, use any two distinct values $\gamma^1, \gamma^2$, to obtain samples $\alpha_1^\ell, \ldots \alpha_M^\ell$ ($\ell=1,2$) from the conditional posterior distribution $\pi(\alpha|\gamma^\ell, \bm y)$. From $\alpha_1^\ell, \ldots \alpha_M^\ell$ ($\ell=1,2$), we can compute fast and accurate approximations of $A$, $B$ and $C$ using the sample mean and variance of each chain. These approximated values can then be used to rapidly speed up the second step of DS, since it will only require taking draws from the appropriate normal distribution. 

It is important to note that this procedure is numerical in spirit, and can be readily generalized. The problem specific assumptions that we assert are (i) normality of the conditional posterior distribution, (ii) constant variance of the conditional posterior distribution and (iii) linear mean of the conditional posterior distribution given $\gamma$. 

In the next section, we will show that the assumption of normality of the conditional posterior distribution can be altered with relative ease, which is convenient for handling any constraints on the support for $\alpha$. Moreover, we have seen in a number of examples that the final approximation of the cut-distribution is often robust to misspecification of the conditional posterior distribution. The second and third assumptions may not be reasonable in more complicated problems, but these assumptions can be weakened considerably by using a more sophisticated approach (e.g., Gaussian process regression) to learn the posterior mean and variance of $\alpha$, conditional on each sample $\gamma$.

\subsection{The Univariate ECP Algorithm}\label{sec:uv_ecp}

\Cref{sec:speedingup} presents an example where $\bm \gamma$ and the first two moments of the conditional posterior distribution $\pi(\bm \alpha|\bm \gamma, \bm y)$ have a simple linear relationship. That example naturally suggests a broader and more general approach for sampling from the cut-distribution; employing a statistical model to link $\bm \gamma$ with key quantities of $\pi(\bm \alpha|\bm \gamma, \bm y)$ such as parameters or sufficient statistics. In this section, we explore an algorithm for sampling from the cut-posterior which does just that. 

Assume that we are given data $\bm y$, and a model which relates $\alpha \in A \subseteq \mathbb R$, $\bm\gamma \in \Gamma \subseteq \mathbb R^q$ and $\bm y$. 
The setting this methodology addresses is where $\bm\gamma | \bm z$ can be sampled efficiently and that sampling $\alpha$ from its full conditional posterior distribution is time-intensive (requiring MCMC). Define $L$ to be the computational budget, which describes the number of different $\bm\gamma$ values for which $\alpha$ will be sampled from its full conditional posterior distribution $\pi(\alpha | \bm\gamma, \bm y)$. Our goal is to construct an algorithm which produces a better approximation of the cut-distribution than DS given the same computational budget. 

In this section we assume that there is a single parameter of interest $\alpha$ and in \cref{sec:mv_ecp} we will generalize the approach to multivariate $\bm \alpha$. The success of the ECP algorithm requires the following set of assumptions. 
\begin{assumption} \label{ass:1} For fixed $\bm y$ and $\bm\gamma \in \Gamma$, the conditional posterior distribution $\pi(\alpha|\bm\gamma, \bm y)$ belongs to a parametric family $\mathcal F$ with parameters $\bm \psi(\bm\gamma) = (\psi_1(\bm\gamma),\ldots,\psi_r(\bm\gamma))$, where each $\psi_j(\cdot)$ is a continuous function of its inputs.
\qed
\end{assumption}

Assumption 1 is strict and generally not true in practice, however there is often a parametric family which provides a good approximation to the conditional posterior distribution. In many cases, selecting a normal distribution for $\mathcal F$ $($with $\psi_1(\bm \gamma) = \mu(\bm\gamma)$ and $\psi_2(\bm \gamma) = \sigma(\bm\gamma))$ is a reasonable default, due to limit results such as the Bernestein von-Mises theorem \cite{Freedman1963}. When the support of $\alpha$ is bounded, other distributions may be preferred.  Although any parametric distribution is theoretically allowable as a choice for $\mathcal F$, we will constrain ourselves to distributions which satisfy the following assumptions. 

\begin{assumption} \label{ass:2}  The parametric family $\mathcal F$ has the following properties
\begin{itemize}
\setlength\itemsep{-0.5em}
\item[i)] it is easy and fast to sample from $\mathcal F$ conditional on the parameters $\psi_1, \ldots \psi_r$,
\item[ii)] given a set of i.i.d.\@ samples from $\mathcal F$, consistent estimators $\hat{\psi}_1, \ldots, \hat{\psi}_r$ of the parameters $\psi_1, \ldots, \psi_r$ are available.
\qed
\end{itemize}
\end{assumption}

If assumptions 1 and 2 hold, and $\psi_j(\bm\gamma)$ is known for all $\bm\gamma \in \Gamma$, we can sample efficiently from the cut-distribution using 
\begin{equation} \label{eq:samples}
\begin{aligned}
\bm\gamma^\ell &\sim \pi(\bm\gamma | \bm z) \\
\alpha^\ell|(\bm\gamma^\ell, \bm y) &\sim \mathcal F(\alpha |\psi_1(\bm\gamma^\ell), \ldots \psi_r(\bm\gamma^\ell)).
\end{aligned}
\end{equation}
In practice, the functional form of the parameters $\psi_j(\cdot)$ is unknown and must be estimated. This is an important idea of the ECP algorithm, that the computational budget can be spent learning the structure of each $\psi_j(\cdot)$. In the example of \cref{sec:speedingup}, the parameters $\psi_1(\cdot)$ and $\psi_2(\cdot)$ were linear functions of $\gamma$. In non-trivial problems, this structure can be learned by using a more flexible statistical emulator. For this work we limit our emulator choice to those which satisfy our final assumption. 




\begin{assumption}\label{ass:3} 
For every $\epsilon > 0$ and every compact set $\Gamma_c\subseteq\Gamma$, there exists a set $G=\{\bm\gamma^1, \bm\gamma^2, \ldots \bm\gamma^L\} \subseteq \Gamma_c$, and an emulator $\Psi_j(\cdot)$ trained on this set such that
$$\sup_{\bm\gamma \in \Gamma_c}|\Psi_j(\bm\gamma) - \psi_j(\bm\gamma)| < \epsilon. $$
\end{assumption}

Assumption 3 represents a simple idea; as training data becomes increasingly dense in $\Gamma_c$, the prediction error on $\Gamma_c$ becomes arbitrarily small. There are many statistical emulators which can be chosen to satisfy this assumption. Specifically, any emulator $\Psi_j(\bm\gamma)$ that is a continuous function of $\bm\gamma$ and interpolates (e.g., $\Psi_j(\bm\gamma^k) = \psi_j(\bm\gamma^k)$ for every $\bm\gamma^k \in G$) satisfies assumption 3. This is proved in section 1 of the supplementary material.
In our our examples, we use a Gaussian Process (GP) emulator with the squared exponential kernel to estimate each $\psi_j(\cdot)$.

One can think of DS as defining a mixture representation of the cut-distribution where each mixture component is the conditional posterior of $\bm \alpha | \bm \gamma, \bm y$ for some specific $\bm \gamma$. ECP can be thought of similarly. The mixture defined by DS is limited by the computational budget $L$, which defines the number of mixture components. ECP on the other hand, takes the observed finite mixture of size $L$, and uses an emulator to learn an infinite mixture representation of the cut-distribution. The assumptions above ensure that the infinite mixture is a good approximation to the true cut-distribution for large enough $L$.

To summarize, assumption 1 ensures that $\mathcal F$ is properly specified for each mixture component so that if the true $\bm \psi(\bm \gamma)$ is known, $\mathcal F$ is equal to the conditional posterior of $\alpha$ given $\bm \gamma$. Assumption 2 guarantees that the parameters of $\mathcal F$ are consistently estimated for each mixture component. Assumption 3 guarantees that predicted parameters are correct to arbitrary precision given a sufficiently dense training set. We have found that ECP is fairly robust to assumption 1 and many choices of $\mathcal F$ can lead to a very good approximation. For example, if $\mathcal F$ is chosen to be Gaussian, the finite mixture can approximate $\pi_\text{cut}$ to arbitrary precision as Gaussian mixtures are universal density approximators \citep{goodfellow2016deep}.

Given $L$ samples of $\bm \gamma$ are used to train the emulators, $\Psi_1(\cdot),\ldots,\Psi_r(\cdot)$, the sampling procedure in \cref{eq:samples} becomes 
\begin{equation} \label{eq:samples_ecp}
\begin{aligned}
\bm\gamma &\sim \pi(\bm\gamma | \bm z) \\
\alpha|(\bm\gamma, \bm y) &\sim \mathcal F(\alpha |\Psi_1(\bm\gamma), \ldots \Psi_r(\bm\gamma)).
\end{aligned}
\end{equation}

After taking $M$ samples $\bm \gamma^1,\ldots,\bm \gamma^M \sim \pi(\bm \gamma | \bm z)$ the cut-distribution's empirical density function is given by 
\begin{equation}
    \pi_\text{cut}^M(\alpha | \bm y) = \frac{1}{M}\sum_{s=1}^M \mathcal F\left(\alpha | \Psi_1(\bm\gamma^s), \ldots \Psi_r(\bm\gamma^s)\right).
\end{equation}

This is only an approximation of the desired distribution but we will demonstrate shortly that the approximation is typically much more accurate than DS given similar computation time. Furthermore, this approximation converges to the true cut-distribution when assumptions 1-3 hold and $\Gamma$ is compact. This is specified in theorem 1 and proved in section 1 of the supplementary material. 

\begin{theorem}
If assumptions 1-3 hold, $\Gamma$ is compact, and the distribution function $F_{\alpha}$ of $\mathcal F$ is continuous with respect to $\psi_1,\ldots,\psi_r$ $\forall \; \alpha$, then $\pi_{\text{cut}}^M \xrightarrow{p} \pi_{\text{cut}}$ as $m,M,L\rightarrow\infty$ if $\{\bm\gamma^1,\ldots,\bm\gamma^L\}$, and $\{\bm\gamma^1,\ldots,\bm\gamma^M\}$ are i.i.d. samples from $\pi(\bm\gamma|\bm z)$.
\end{theorem}

For cut parameter sets of small to moderate dimension, the ECP algorithm thrives, but it may struggle in higher dimensions since $L=\mathcal O(2^q)$ MCMC procedures should be budgeted for reliable approximation. Although no recommendation can work for all problems, we have found $L=2^q+4q+1$ to be a sensible default in our work; it scales with the number of vertices in the hypercube but also guarantees a center point and some additional points along each axis. The ECP algorithm is summarized in \cref{alg:ECP} in it's general form ($A \subseteq \mathbb R^p$). 

Note that if $\mathcal F$ is chosen to be a normal distribution, then sampling from the conditional posterior distribution (\cref{alg:ECP}, line 11) and using sample based estimators for $\psi_1,\ldots,\psi_r$ (line 12) can be replaced by the Laplace approximation \citep{laplace_approx}. The Laplace approximation has the potential to result in significant computational savings by avoiding costly MCMC. Results for the example presented in \cref{sec:examples} 
using the Laplace approximation are given in section 2 of the supplementary material.

\begin{algorithm}[!ht]
    \renewcommand{\baselinestretch}{1.15}
    \selectfont
    \footnotesize
	\caption{Sampling from the cut-distribution via the ECP algorithm}
	\label{alg:ECP}
	\begin{algorithmic}[1]
		\Statex \Require{\\
		\quad$L$ - the budget \\
		\quad$M$ - number of samples of $\bm \gamma$ used for the final approximation\\
        \quad$m$ - number of samples of $\bm \alpha$ drawn for each $\bm \gamma$ \\
		\quad$\pi$ - the distribution $\pi(\bm\gamma|\bm z)$ \\
		\quad$\mathcal F$ - distributional assumption for conditional posterior distribution. \\
		\quad$\bm y$ - data }
		\Statex
		\Function{\texttt{ECP-SAMPLE}}{$L, M, \pi, \mathcal F, \bm y$}

		\Statex \# Phase 1: Sampling from $\pi(\bm\gamma|\bm z)$ and estimating parameters for $\mathcal F$
		\State $(\bm\gamma^1, \bm\gamma^2, \ldots, \bm\gamma^L) \sim \pi$
        \Comment{i.e., using MCMC or Support Points (\cref{sec:DS_DOE})}
                
		\For{$\ell = 1$ \text{ to } $L$}
			\State $(\bm \alpha_\ell^1, \bm \alpha_\ell^2, \ldots \bm \alpha_\ell^m) \sim \pi(\bm \alpha | \bm\gamma^\ell, \bm y)$ 
            \Comment{Sample from conditional posterior using MCMC}
			\State$(\hat\psi_1^\ell, \hat\psi_2^\ell, \ldots \hat\psi_r^\ell) \leftarrow \texttt{estimate-parameters}(\{\bm \alpha_\ell\}_{s=1}^m, \mathcal F)$
            \Comment{Estimate parameters for GP}
		\EndFor
		
		\Statex \# Phase 2: Training the Gaussian Process and drawing final samples from $\mathcal F$
		\For{$j = 1$ \text{ to } $r$} 
			\State $\Psi_j(\cdot) \leftarrow \texttt{train-GP}((\bm\gamma^1, \bm\gamma^2, \ldots \bm\gamma^L), (\hat\psi_j^1, \hat\psi_j^2, \ldots \hat\psi_j^L))$
            \Comment{Train the Gaussian Process for each parameter $\psi_j$}
		\EndFor
		
		\For{$i = 1$ \text{ to } $M$} 
			\State $\bm\gamma^i \sim \pi$ 
            \Comment{Sample from $\pi(\bm \gamma | \bm z)$}
			\State $\{\bm \alpha^i\} \sim \mathcal F\left(\alpha|\Psi_1(\bm\gamma^i), \ldots \Psi_r(\bm\gamma^i)\right)$ 
            \Comment{draws from the conditional posterior approximation}
		\EndFor
		
		\State \Return{($\{\bm \alpha^1\},\{\bm \alpha^2\}, \ldots,\{\bm \alpha^M\}$)}
		\EndFunction
	\end{algorithmic}
\end{algorithm}



\subsection{Diamond in a Box, Revisited}\label{sec:DB_analysis}

We return to the Diamond in a Box example to illustrate the ability of ECP to accurately represent the true cut-posterior distribution using a very small budget. We show empirically that DS does not represent the true posterior as well as ECP when the budget is limited. These methods are also compared to DS with Little's aggregation \cite{little1992}, where a normal distribution is fit to the $mL$ samples from DS. Samples from this normal distribution are used to approximate the cut-distribution. For clarity we will refer to DS with Little's aggregation as DS with a normal approximation. If the cut-distribution is approximately normal, this has the potential to improve the accuracy of the approximation. This example is especially relevant for illustrating the accuracy of each method given that the true cut-posterior $\pi_\text{cut}(\alpha| \bm y)$ is known.

Suppose that the true weight of the diamond is 1 gram and is weighed $n_1=10$ times. The box is 10 grams and similar boxes are weighted $n_p=1000$ times. Those data are used to define $\gamma|\bm z \sim N(10,0.1^2)$. The box with the diamond inside is weighed $n_2=100$ times. The scale has a known variance of $\sigma^2=.01$. Lastly, define an informative prior distribution for the true diamond weight $\alpha \sim N(1,0.1^2)$ representing a 10\% error around the truth.

We consider the ability to recover the true cut-posterior for $\alpha$ using DS, DS with a normal approximation, and ECP with $\mathcal F$ a univariate normal distribution under varied budgets $L$. For DS with a normal approximation, we simply fit a normal distribution to the samples of $\alpha$ conditional on $\gamma$ obtained from DS. Similar to ECP, an important benefit of a normal approximation is that a large number of samples of $\alpha$ can be obtained from the fitted distribution. In this case, the true cut-posterior follows a normal distribution, so we expect DS with a normal approximation to do very well, and indeed it does provide a significant improvement over DS for small $L$.

\begin{figure*}
    \centering
    \includegraphics[width=\linewidth]{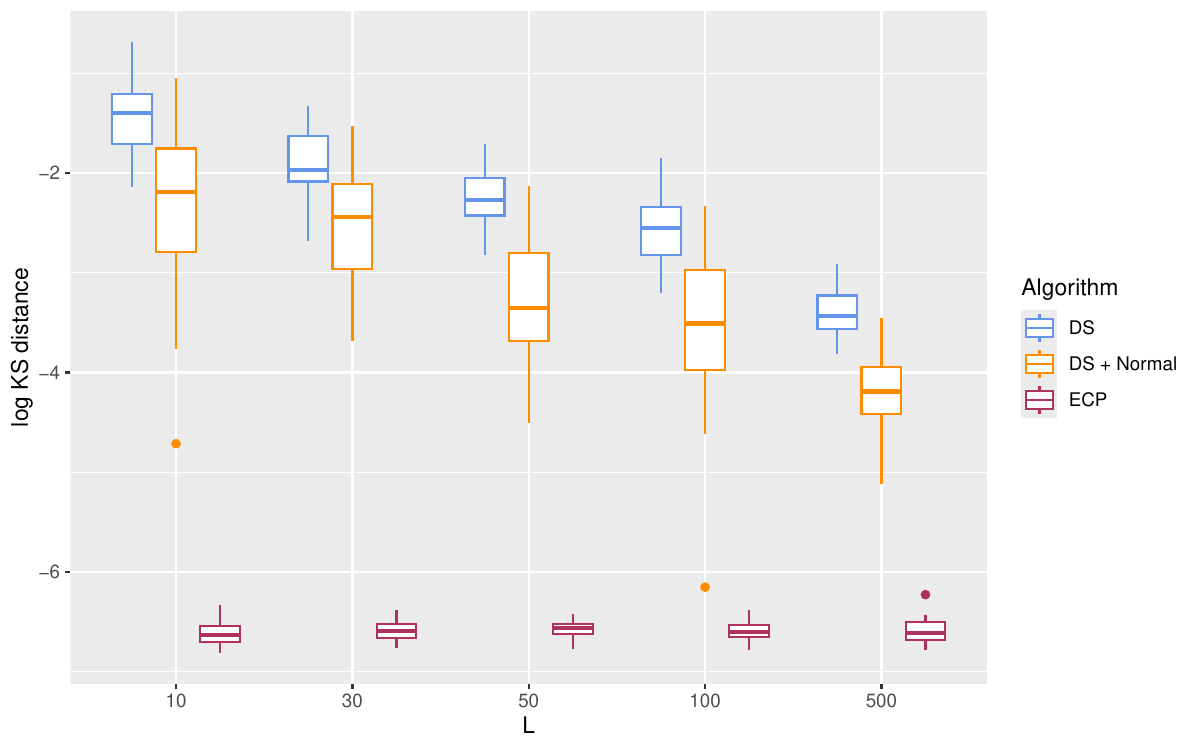}
    \caption{Logarithm of the KS distance for the Diamond in a Box cut-posterior. DS, DS with a normal approximation, and ECP are compared over a range of budgets $L$. ECP provides significantly better accuracy than DS for all budgets. Boxplots are in the order that they appear in the legend.}
    \label{fig:DIB_D_Stat_random}
\end{figure*}

\Cref{fig:DIB_D_Stat_random} shows the logarithm of the Kolomogorov-Smirnov (KS) distance for each method over 25 sets of i.i.d.\@ samples from $\pi(\gamma|\bm z)$. The KS distance is defined as $D_n = \sup_{\bm\alpha\in A} |F_n(\bm\alpha) - F(\bm\alpha)|$. Here, $F_n(\bm\alpha)$ is the empirical CDF estimated from a sample of size $n$ and $F(\bm\alpha)$ is the true CDF of $\bm\alpha | \bm y$. A smaller distance indicates a better fit of the samples to the distribution. We use the KS distance for it's simplicity and interpretability.

To standardize the methods, 10000 samples are collected from $\pi_\text{cut}(\alpha| \bm y)$ for each, regardless of the budget. For DS, that means setting $m=10000/L$. These large values of $m$ can have negative effects on the approximation due to sample dependence. 

ECP provides a dramatic improvement in the median KS distance for all budgets, indicating a much better fit to the true cut-posterior. Furthermore, ECP seems to have nearly optimal performance even for a very small budget of $L=10$. When the budget is increased, DS and DS with a normal approximation improve, but are still not competitive with ECP even at a $L=500$. We would not expect DS with a normal approximation to do well for cut-posteriors that deviate significantly from a normal distribution, such as multimodal distributions. The DS approach will approach ECP in accuracy as the budget increases, but even for this very simple problem, that may require many thousands of imputations.

The DS approach with or without a normal approximation is also much more variable than ECP. Variability is mostly due to sampling variability in $\gamma | \bm z$, with minor contributions from the sampling of $\alpha | \gamma, \bm y$. We will discuss this in more detail in \cref{sec:DS_DOE}, and discuss improvements to the DS algorithm which reduce variability and increase accuracy. 

ECP also has variability due to sampling variability in $\gamma$, although that variability is very small for the DB problem. We have generally found ECP to be quite robust to the sample set of $\gamma$, so long as they adequately cover the tails of the distribution. Issues can arise when the training samples do not cover the tails of the distribution so that predictions are sometimes made in extrapolatory regimes. One method we have found to be successful in cases where the training sample set does not cover the tails, is to inflate the variance of the $\gamma$ samples. If the variance is inflated so that the tails are adequately captured there is little risk of extrapolation when making predictions. See section 4 of the supplement for an approach to variance inflation when $\Gamma$ is a bounded space. 


This example serves to illustrate the accuracy of ECP and the improvement over DS, especially when the budget is small.

\subsection{A Multivariate ECP Algorithm}\label{sec:mv_ecp}
The univariate ECP algorithm can be naturally extended to the case where $\bm\alpha \in A \subseteq \mathbb R^p$. The unifying feature is the parametric distribution $\mathcal F$ which is specified for the conditional posterior distribution $\pi(\bm \alpha | \bm\gamma, \bm y)$. Although any distribution with the aforementioned properties can be a valid choice for $\mathcal F$, we will limit our discussion to the multivariate normal distribution. In particular, we assume that $\pi(\bm\alpha|\bm\gamma,\bm y)$ is multivariate Gaussian with
\begin{equation}
\label{eq:MVN}
\begin{aligned}
&\alpha_i \sim N\left(\mu_i(\bm\gamma), \sigma^2_i(\bm\gamma)\right), \ i=1,2,\ldots p \\[1.2ex]
&\text{Cov}(\alpha_i, \alpha_j) = \sigma_{ij}(\bm\gamma), \ 1 \leq i < j \leq p.
\end{aligned}
\end{equation}
This implies that there are $r = 2p + \binom{p}{2}$ total $\psi_j(\cdot)$ functions which must be estimated. When the dimension of $\bm\alpha$ is small, i.e. $p=2,3,4$, then this is feasible. For larger sets of physical parameters, a pre-posterior sensitivity study may be useful for obtaining a sparse set of parameters \cite{arendt2016, friedman2008}.

The algorithm presented in \cref{sec:uv_ecp} which is naturally extended here for $p>1$ has the potential issue that predicted covariance matrices are not guaranteed to be positive-semi-definite because emulators are fit independently to each $\sigma_{ij}(\bm\gamma)$ \citep{GPalphagamma2023}. If the predicted matrix is not positive semi-definite it must be adjusted, and that adjustment could negatively impact the accuracy of the approximation to the cut-posterior. Throughout this work, the method described in \cite{HIGHAM1988103} is used when necessary to find the nearest positive semi-definite matrix to the predicted covariance matrix. We have not seen any applications where this has noticeably impacted approximation accuracy. We understand that this may not be the case for all applications, especially when parameters $\bm \alpha$ are highly correlated. In such cases, an altered algorithm may be needed, such as using a regression method on covariance matrices which preserves positive semi-definiteness \citep{covregression}.

\subsection{Bootstrapping the Parameters}
One source of error in the ECP approximation comes from Assumption $2$\textit{ii}, which says that the parameters of $\mathcal F$ can be estimated consistently and quickly. In many cases, especially those where we can afford a large $m$, this assumption is reasonable and the corresponding error will be negligible. In instances where this source of uncertainty cannot be ignored, we can propagate the uncertainty forward via the non-parametric bootstrap algorithm \citep{efron1982}. Specifically, rather than associate a single estimate $\hat\psi_j^\ell$ with each $\bm\gamma^\ell$, we use bootstrap resampling to obtain a set of $B$ estimates $\hat\psi_j^{\ell, b}$ $(b=1, \ldots, B)$ associating each estimate with $\bm\gamma^\ell$ in the data set used to train the emulator. If estimating parameters is fast (as is the case for the distributions we have discussed here), obtaining these bootstrap estimates will not require substantial time. The training data for each emulator, however, does grow from $L$ to $LB$, hence the cost of training the emulators (and thus, the cost of ECP) grows by a factor of $\mathcal O(B^3)$ if a GP is used. In this case, alternatives to the full Gaussian process should be considered \citep{liu2020gaussian, BASS_package, bart}. For instance, the heteroskedastic GP approach of \cite{binois2021hetgp} can take advantage of this Kronecker structure, effectively eliminating the $\mathcal O(B^3)$ penalty. 

\subsection{Complexity Analysis}

We define the computational complexity of the ECP algorithm under the assumption that sampling from the prior distribution is fast, and sampling from the conditional posterior distribution is time consuming. 

Let $T$ denote the cost to run a single MCMC chain targeting $\pi(\bm \alpha | \bm \gamma, \bm y)$. 
For a fixed budget $L$ the cost of DS is $\mathcal{O}(LT)$ and if $P\leq L$ parallel processors are used we have $\mathcal{O}(\frac{L}{P}T)$. The ECP algorithm requires fitting $r$ emulators along with the $L$ MCMC chains. If a GP is used, this gives a total complexity of $\mathcal{O}(\frac{L}{P}T + \frac{r}{P}L^3)$ when $P \leq min\{r,L\}$ parallel processors are used.

Complexity analysis is important for an honest comparison of the approaches. For instance, it suggests that ECP with standard GP's may not be feasible for $L > 1000$, suggesting an upper bound on the number of allowable parameters $\bm \gamma$ (somewhere around q = 10 based on our recommendation that  $L \geq 2^q + 4q+ 1$). For larger $L$ there are a number of scalable GP implementations which could be used (e.g., see \cite{sVecchia} and references therein). In fact, any regressor could be used and there are many that satisfy assumption 3. We caution that complexity analysis can also be misleading here. The ECP algorithm is expected to take slightly longer than DS for the same budget. However, the examples of \cref{sec:DB_analysis} and \cref{sec:examples} illustrate that ECP can provide a much more accurate approximate to the cut-distribution for the same (and often smaller) budget $L$.

\section{Direct Sampling with Design of Experiments}\label{sec:DS_DOE}

In \cref{sec:DB_analysis} we saw that the DS can lead to an inaccurate approximation of the cut-posterior with small computational budgets. A reasonable question to ask is, can DS be improved using DOE methods? In this section we show that indeed it can. If we assume that the cut-posterior is sensitive to $\bm \gamma$, the samples $\{\bm\gamma_1,\ldots,\bm\gamma_L\}$ that are used to evaluate the conditional posterior distribution $\pi(\bm \alpha | \bm \gamma, \bm y)$ are critical to the performance of DS. That is, the accuracy of inference on the cut-distribution is dependent on how well the samples represent the true $\pi(\bm \gamma|\bm z)$. 
This is intuitive when examining \cref{eq:cutdistr}, as the cut-distribution is defined as an integral over $\pi(\bm \gamma|\bm z)$. DOE methods that provide improved integration error compared to random sampling are directly applicable to the approximation of the integral in \cref{eq:cutdistr}.


\subsection{Sampling from the Prior Distribution}

We seek a set of $L$  points, $\{\bm\gamma_1,\ldots,\bm\gamma_L\}$ which represent the distribution $\pi(\bm \gamma|\bm z)$ well. When $L$ is large, random sampling and MCMC are likely to accomplish this, but when $L$ is small to moderate that may not be the case and so extra care must be taken.

Consider the Diamond in a Box problem from \cref{sec:diamond} once again. Since the distribution $\pi(\gamma | \bm z)$ is known, i.i.d.\@ samples can be generated. However, we demonstrate that random sampling with small $L$ often results in sets of samples which poorly reflect the true distribution due to sampling variability. We show that this issue can be easily remedied by comparing random sampling to three more sophisticated sampling methods over a range of budgets. Latin hypercube sampling (LHS) \citep{mckay1992latin} provides a space filling design with respect to the known distribution. We find this method to be very effective in generating representative samples, but is limited by the need for a quantile function, and can be difficult in high dimensions when correlations between inputs exist. Work by \cite{helton2003latin} demonstrates that LHS provides a more stable estimate of a distribution compared to random sampling. Minimum energy designs \citep{joseph2015mined} (MinED) and Support Points \citep{mak2018support} (SP) are methods which select samples that minimize an energy criterion. The energy criterion balances a space filling property with a desire to place points in high density regions. MinED can be applied when the quantile function is not available, requiring only the unnormalized density of $\gamma|\bm z$. SP requires either a distribution function, or a large set of samples, usually generated from MCMC. The set of $L$ support points for a given sample is the subsample of size $L$ that minimizes the energy criterion. It is demonstrated by \cite{mak2018support} that support points can provide a substantial improvement compared to random sampling and a QMC method. 

These three approaches are compared to i.i.d.\@ random sampling in their ability to generate sample sets which represent the true distribution, $\gamma \sim N(\mu_{\gamma},\sigma_{\gamma})$. The approaches are judged both qualitatively though density estimates and quantitatively using the KS distance. The KS distance is used here for the same reasons discussed in \cref{sec:DB_analysis}. A relatively small budget of $L=30$ is used to sufficiently illustrate the issues with random sampling. 

To account for sampling variability in the comparison of methods, each procedure is repeated 25 times. For random sampling this simply means generating a new set of $L$ samples with a different random seed. For LHS, the R package \texttt{lhs} \cite{carnell2016package} is used to generate a different minimax LHS sample on $[0,1]$ which is passed through the true quantile function for each repetition. MinED uses those minimax designs as starting points, and adjusts them according to the energy criterion. Support points are generated based on the fully-known target distribution.

\begin{figure*}[hb!]
    \centering
    \begin{subfigure}[t]{0.32\textwidth}  
        \centering
        \includegraphics[width=\textwidth]{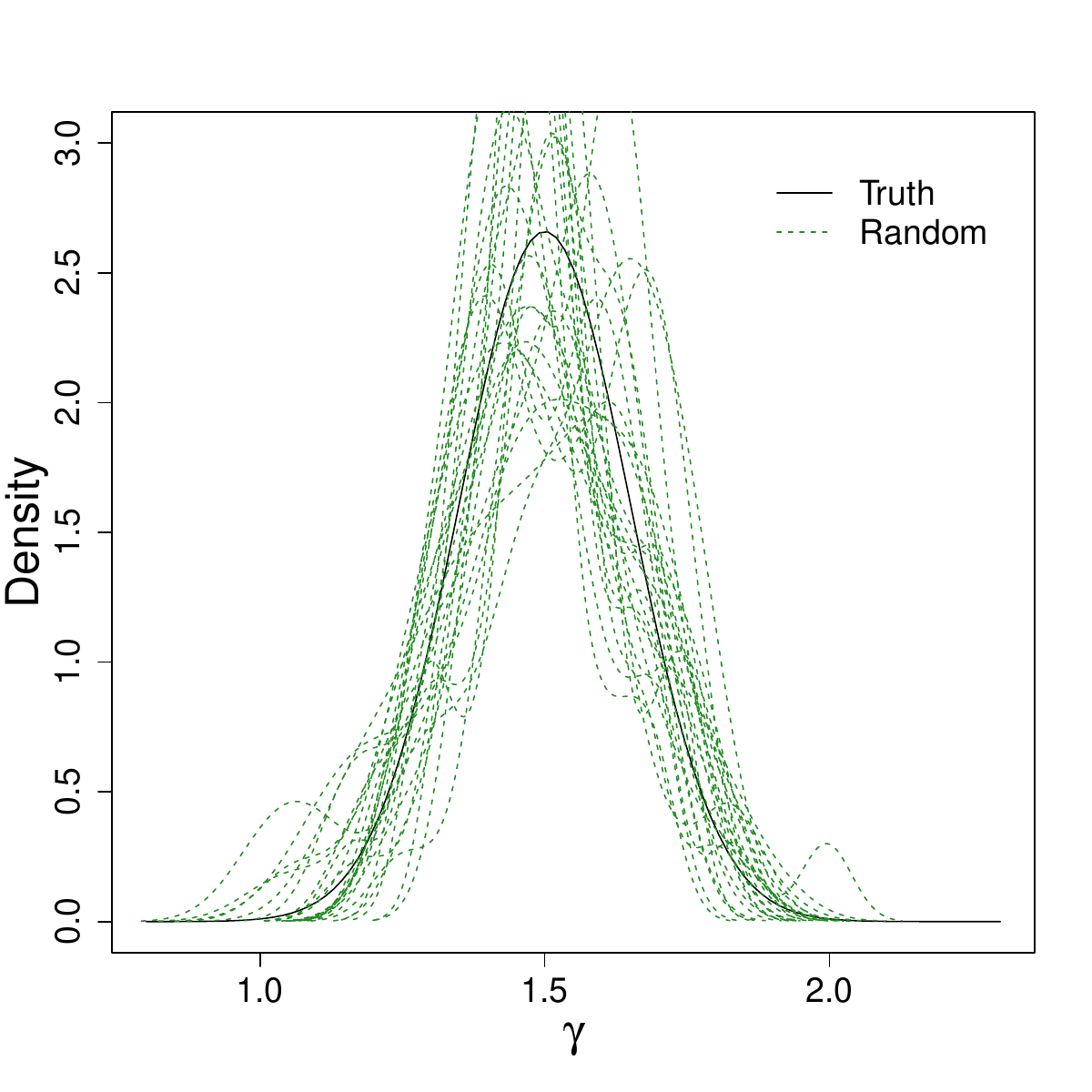}
        \caption{Kernel density estimates using random sampling.}
        \label{fig:rand_approx_DIB}
    \end{subfigure}
    \hfill
    \begin{subfigure}[t]{0.32\textwidth}  
        \centering 
        \includegraphics[width=\textwidth]{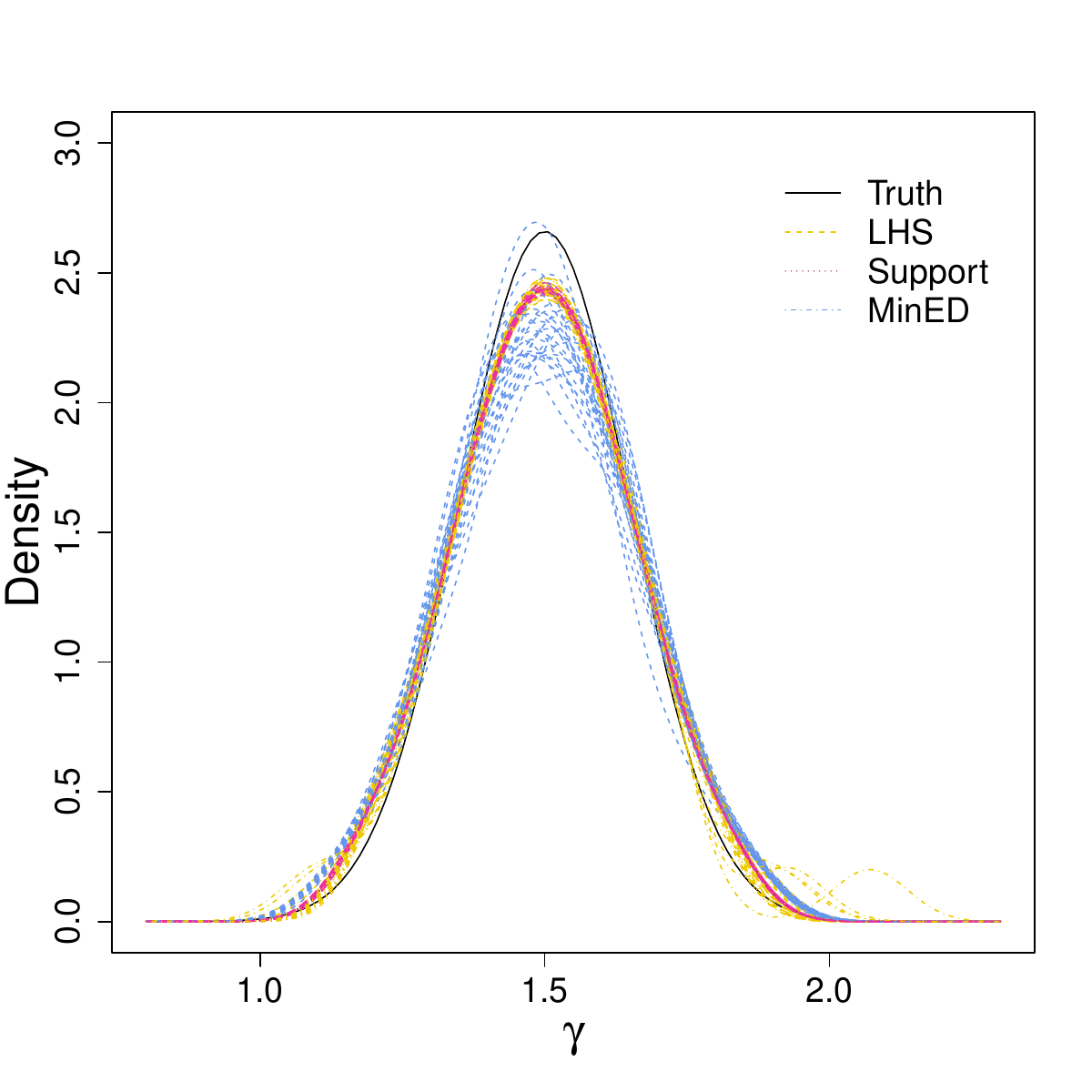}
        \caption{Kernel density estimates using DOE sampling approaches.} 
        \label{fig:DOE_approx_DIB}
    \end{subfigure}
    \hfill
    \begin{subfigure}[t]{0.32\textwidth}  
        \centering 
        \includegraphics[width=\textwidth]{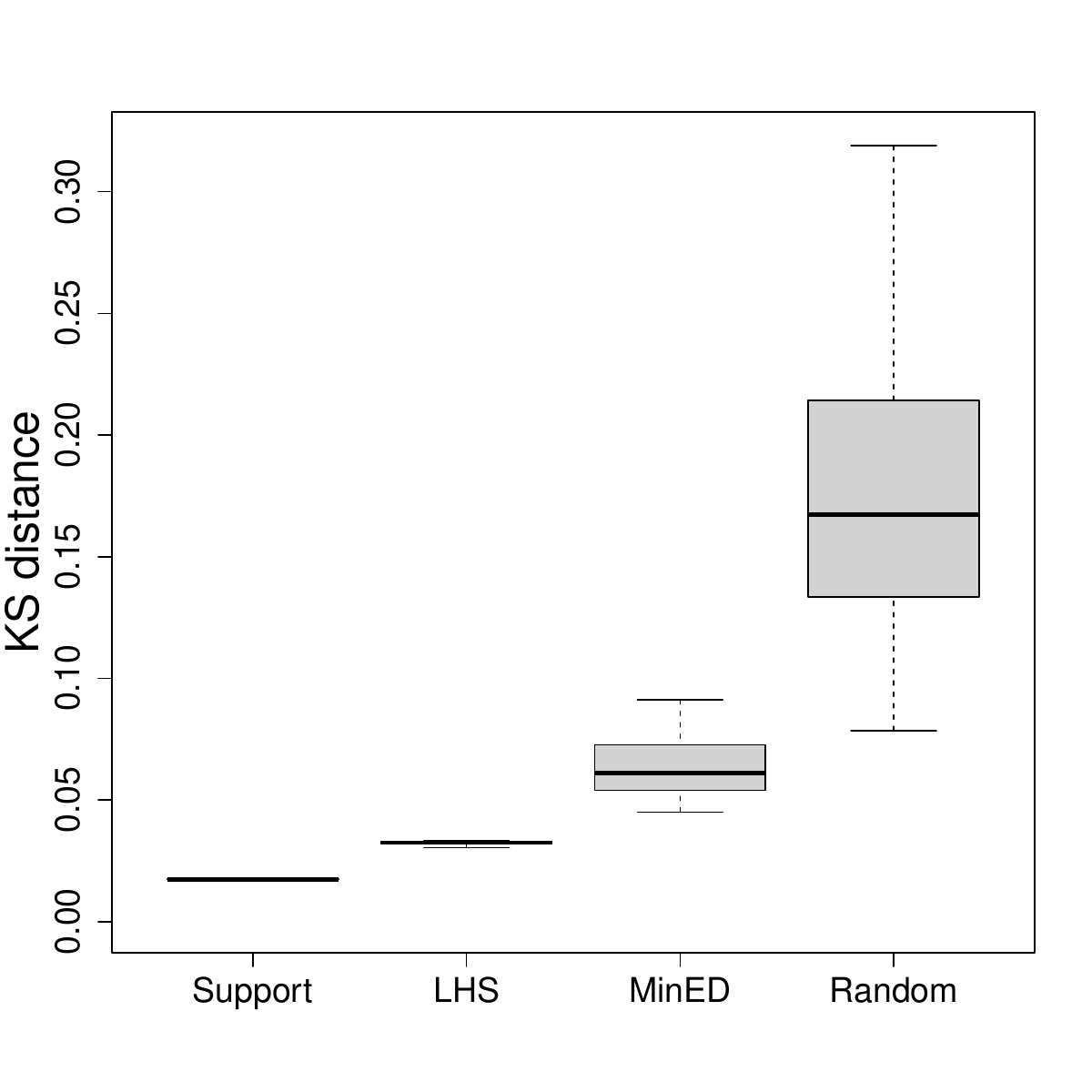}
        \caption{Approximation accuracy (KS distance).} 
        \label{fig:KS_approx_DIB}
    \end{subfigure}
    \caption{Approximating $\pi(\gamma|\bm z)$ using a small set of $L=30$ samples. Panel (a) shows density estimates based on 25 sets of random samples. Panel (b) shows density estimates based on samples attained using some DOE approaches. Panel (c) quantifies how well these sample sets approximate $\pi(\gamma|\bm z)$ using KS distance.}
    \label{fig:approx_DIB_dist}
\end{figure*}

\Cref{fig:rand_approx_DIB} shows qualitatively that random sampling is highly variable in its outcomes and that these random draws are likely to misrepresent the distribution. \Cref{fig:DOE_approx_DIB} shows that the DOE approaches more accurately represent the true distribution with less variability. Among the DOE approaches, MinED is more variable than LHS and SP, and tends to be slightly less accurate. We have seen similar behavior with other examples and therefore only recommend the use of MinED when LHS or SP are infeasible. 
Boxplots of the KS distance over the 25 runs of the experiment are given in \cref{fig:KS_approx_DIB}. These results clearly show that the DOE methods provide a more accurate and less variable representation of the distribution. For this problem and others, we have observed that SP tends to provide the best accuracy among the chosen methods, and is a significant improvement over random sampling. 

While we have found SP to generally be an improvement over MinED, it is important to note that the SP algorithm requires the distribution function or a large set of samples to optimize over, limiting its applicability to situations where such samples are readily available. MinED on the other hand, can be applied using only the unnormalized density. In the remaining examples presented in this work, those samples are available and so SP is used rather than MinED.

\subsection{Approximating the Cut-Distribution}

The accuracy, or lack there of, of the approximation to $\pi(\gamma|\bm z)$, will determine the accuracy of the resulting approximation to the cut-distribution when using the DS approach. In \cref{sec:DB_analysis} we applied the DS algorithm using the random sample sets shown in \cref{fig:rand_approx_DIB}. Here, we apply the DS approach using each of the DOE sample sets that generate the distributions shown in \cref{fig:DOE_approx_DIB}. We also apply the ECP algorithm with these DOE approaches to show that ECP is less sensitive to these samples.

\begin{figure*}
    \centering
    \includegraphics[width=\linewidth]{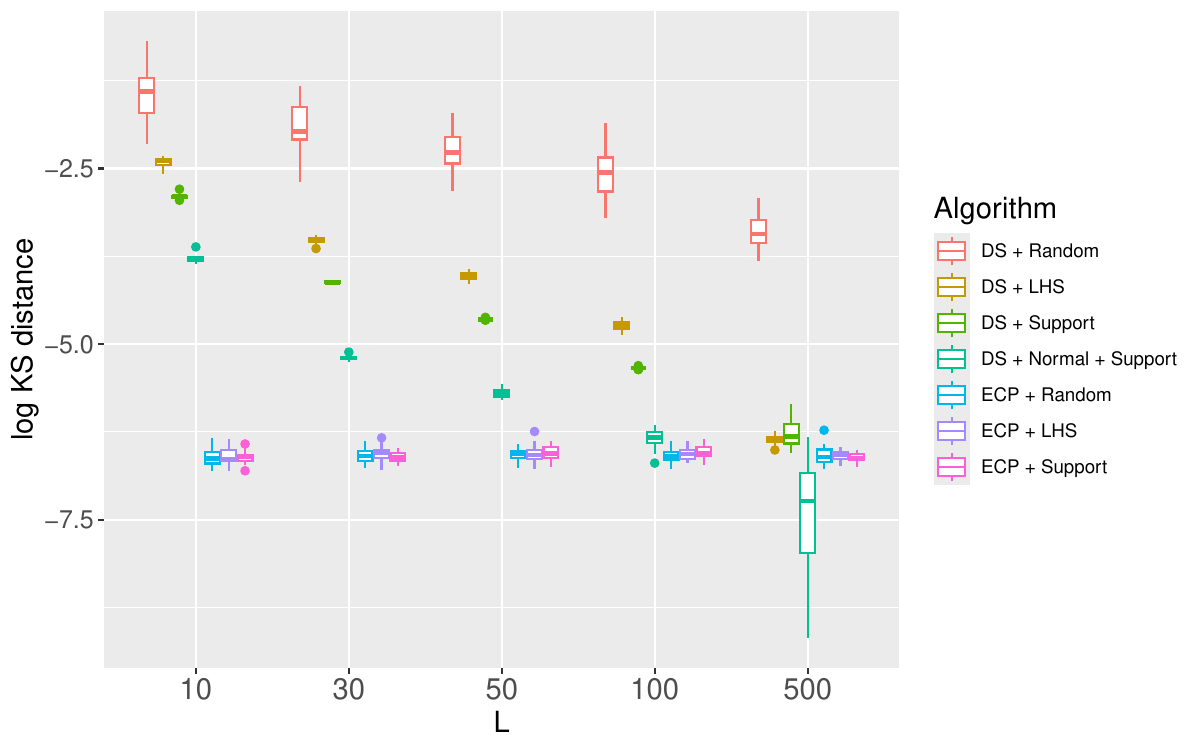}
    \caption{Logarithm of the KS distance between the approximated $\pi_\text{cut}(\alpha | \bm y)$ and the truth for the DB example. Approximation accuracy is compared over a range of sampling techniques for the prior distribution $\pi(\gamma | \bm z)$. DS + Normal + Support indicates DS with a normal approximation using support points for selecting the $L$ samples from $\pi(\gamma | \bm z)$. Boxplots are in the order that they appear in the legend.}
    \label{fig:DIB_D_Stat}
\end{figure*}

\Cref{fig:DIB_D_Stat} shows the logarithm of the KS distance between the approximated $\pi_\text{cut}(\alpha | \bm y)$ and the truth for DS and ECP with random sampling, LHS, and SP. We omit MinED for this comparison as Support Points and LHS provide notably better representations of the prior distribution (\cref{fig:KS_approx_DIB}). For all $L$, DS with LHS and SP give far more accurate estimates of $\pi_\text{cut}(\alpha | \bm y)$ than DS with random sampling. SP is preferred to LHS once again. When DS with SP also uses a normal approximation, it is competitive with ECP for large enough budgets ($L \geq 100)$. This is not surprising since the true cut-posterior is normal. If this was not the case, a normal approximation could lead to a worse approximation of the cut-posterior. Without a normal approximation, DS only becomes competitive with ECP for a large budget of $L=500$.

This example clearly illustrates that DOE techniques can dramatically improve the accuracy of the DS algorithm. It also demonstrates that the ECP algorithm is not as sensitive to the sampling method, so long as the samples of $\gamma$ are representative enough to train the emulators (e.g., they cover the tails of the prior distribution).

\section{Ecological Study}\label{sec:examples}

In this section we explore a modified Ecological study from \citep{plummer2015}. We compare 3 methods for sampling from the cut-posterior (i) DS, (ii) DS with a normal approximation and (iii) ECP. For sampling from $\pi(\bm\gamma|\bm z)$ we use SP given the results in \cref{sec:DB_analysis}. In all of our examples, each emulator $\Psi_j(\cdot)$ is taken to be the mean function of a standard Gaussian process, using the \texttt{hetGP::mleHomGP()} function in R \cite{binois2021hetgp}. 

The ecological study discussed in \cite{plummer2015} is intentionally modified to better illustrate the challenges associated with many important practical applications. In the Diamond in a Box problem both the prior distribution and the cut-posterior are known normal distributions, which allows for direct sampling and comparison to the truth. In most real world applications this will not be the case. Here, we present an example which alters an important cut model application from \cite{plummer2015} so that neither of these distributions are available in closed form. Since DS with large $L$ is the gold standard for solving such a problem \citep{plummer2015,jacob2017BetterTogether}, we use the DS posterior with $L=10^4$ as the target solution and assess the ability of ECP and DS to approximate this solution when the budget is much smaller. The study discussed in \cite{plummer2015} is augmented as follows.

In population $j$ ($j=1,\ldots,13$), the outcome $y_j$ is the number of cervical cancer cases arising from $T_j$ woman-years of follow-up; in a (separate) sample of $N_j$ women from the same population, $z_j$ is the number of women infected with high-risk HPV. Parameters $\phi_j$ and $\lambda_j$ represent the high-risk HPV prevalence and the cancer incidence rate (per 100,000) in population $j$ respectively. We assume that $\phi_j$ and $\lambda_j$ are log-linearly related. Unlike in \cite{plummer2015}, however, we assume that a prior distribution for $\phi_j$ cannot be specified directly; rather, prior information for $\phi_j$ is encoded through a series of predictors $\gamma^1, \ldots, \gamma^K$ (for which prior distributions can be specified) in combination with some population-specific constants $C_j^1, C_j^2, \ldots C_j^K$. The full data, $\mathcal D_j = \{T_j, N_j, C_j^1, \ldots C_j^K\}$ $(j=1,\ldots, 13)$, is provided in section 2 of the supplement. The full model can be specified as follows
\begin{equation}
    \begin{aligned}
        y_j|\bm\alpha,\bm\gamma, \mathcal D_j &\stackrel{\text{ind}}{\sim} \text{Poiss}(\lambda_j T_j)  \\[1.1ex]
        z_j|\bm\gamma, \mathcal D_j &\sim \text{Binom}(N_j, \phi_j) \\[1.1ex]
        \log\lambda_j &= \alpha_1 + \alpha_2\phi_j(\bm \gamma, \bm C_j) \\[1.1ex]
        \phi_j(\bm \gamma, \bm C_j) &= g\left(\gamma_1^{-C_j^1}, \ldots, \gamma_k^{-C_j^k}\right), 
    \end{aligned}
\end{equation}
for $j=1,\ldots, 13$ with prior distributions
\begin{equation}
    \begin{aligned}
        \gamma_k &\stackrel{\text{ind}}{\sim} \text{Beta}(2, 2), \ k = 1,\ldots, K \\
        \alpha_i &\sim \mathcal{N}(0,100^2), \ i = 1,2
\end{aligned}
\end{equation}
which are chosen to represent a lack of strong prior knowledge about the parameters. For this particular example, we take $K=5$ and specify a complicated non-linear form for $g$,
\begin{equation}
    \begin{aligned}
    g(x_1, \ldots, x_5) = \frac{19}{700} \Big[
    & 1.35 
    + e^{x_1}\sin \left(13(x_1 - 0.6)^2\right) e^{x_2} \sin(7x_2) \\
    & + \frac{1}{38} \left(x_3 \sqrt{x_4} \sin(2\pi x_5)^2\right) 
    \Big]^{1/3}.
    \end{aligned}
\end{equation}

The cut-posterior is given by $\pi_\text{cut}(\bm\alpha | \bm y) = \int \pi(\bm\alpha | \bm \gamma, \bm y) \pi(\bm\gamma | \bm z)  d\bm\gamma$ where 
\begin{equation}
\begin{aligned}
    \pi(\bm\alpha \mid \bm \gamma, \bm y) &\propto 
    \prod_{j=1}^{13} \text{Poiss}\left(y_j \mid T_j \lambda_j\right) \pi(\bm\alpha) \\ 
    &= \prod_{j=1}^{13} \text{Poiss}\left(y_j \ \Big|\ T_j \exp\left\{\alpha_1 + \alpha_2 
    \phi_j(\bm \gamma, \bm C_j)\right\} \right) \\
    &\qquad \times \mathcal{N}(\alpha_1 \mid 0, 100^2) \mathcal{N}(\alpha_2 \mid 0, 100^2)
\end{aligned}
\end{equation}

and
\begin{equation}\label{eq:gamma_post}
\begin{aligned}
    \pi(\bm\gamma \mid \bm z) &\propto \prod_{j=1}^{13} \text{Binom}\Big(z_j \mid N_j, \phi_j(\bm \gamma, \bm C_j) \Big)  \prod_{i=1}^5 \text{Beta}(\gamma_i \mid 2, 2).
\end{aligned}
\end{equation}

The first step in this example is to generate samples from the conditional posterior distribution in \cref{eq:gamma_post}. An adaptive metropolis algorithm is used to generate a large set of $9\times10^4$ post burn-in posterior samples from which a subsets of $L$ samples can be selected. Marginal posterior density estimates from these samples are shown in section 3 of the supplementary material. 


To simulate ground truth we perform DS with $L=10^4$ and $m=1$ which gives $10^4$ samples from $\pi_\text{cut}(\bm{\alpha}|\bm y)$. We compare to this simulated ground truth by applying DS over a range of budgets $L=10,25,50,100,250,$ and $1000$. For each budget $L$, we generate a set of $L$ support points from the posterior samples of $\bm \gamma | \bm z$ and use those for DS. We standardize the DS approach with different budgets by letting $m=10^5/L$ for each $L$ resulting in $10^5$ total samples approximating the cut-distribution.

To standardize ECP over the different budgets, we predict at the same set of $10^4$ support points derived from the posterior samples of $\bm \gamma | \bm z$ and randomly sample $m=10$ points from each predicted $\mathcal F$. For training the GP emulators, $L$ support points are generated from those same posterior samples of $\bm \gamma | \bm z$. We found that for low budgets ECP benefited from first slightly inflating the variance of the posterior samples of $\bm \gamma | \bm z$ before selecting the $L$ support points. This step ensures there is no extrapolation when predicting from the GP's at new values of $\bm\gamma$ as discussed in \cref{sec:DB_analysis}.

\begin{figure}[t]
    \centering
    \includegraphics[width=1\linewidth]{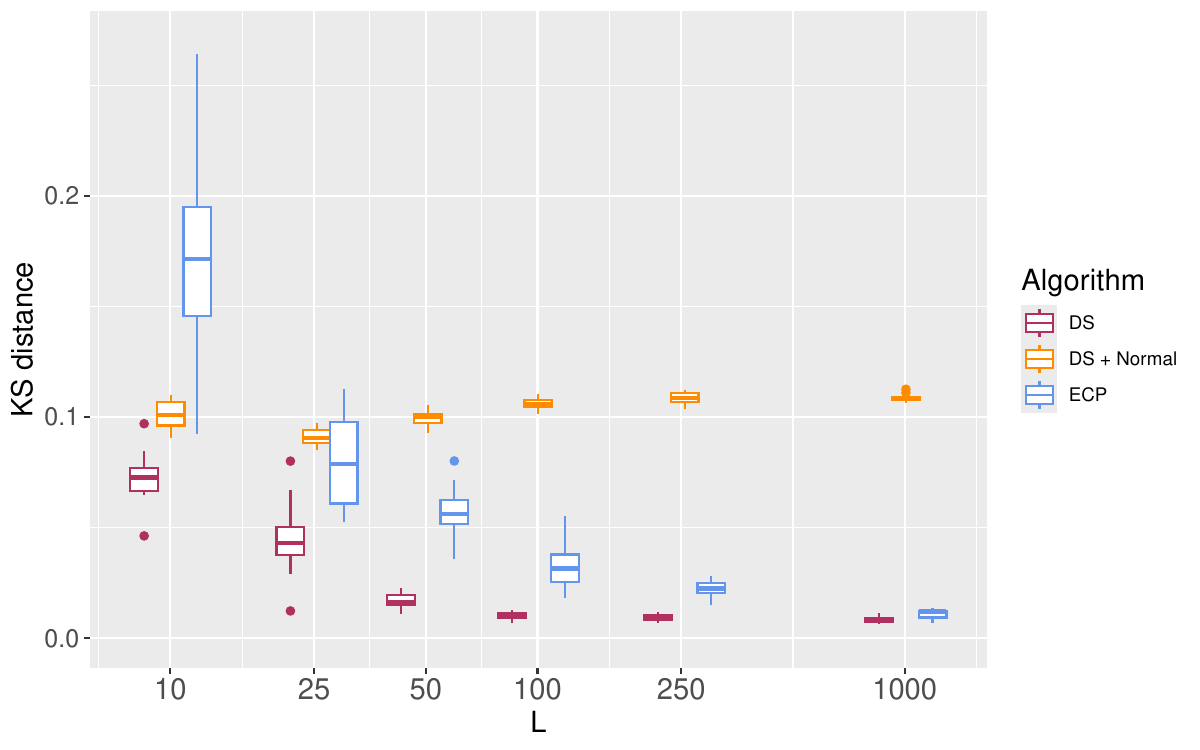}
    \caption{KS distance between the simulated `true' cut-posterior and the estimated cut-posterior using DS, DS with a normal approximation (DS + Normal), and the ECP algorithm. A normal approximation does not improve DS due to non-normal conditional posterior distributions for $\bm\alpha$. The accuracy of ECP is superior to DS for all budgets, and the difference is most significant at low budgets.}
    \label{fig:plummer_KS}
\end{figure}
\begin{figure}[h]
    \centering
    \includegraphics[width=.95\linewidth]{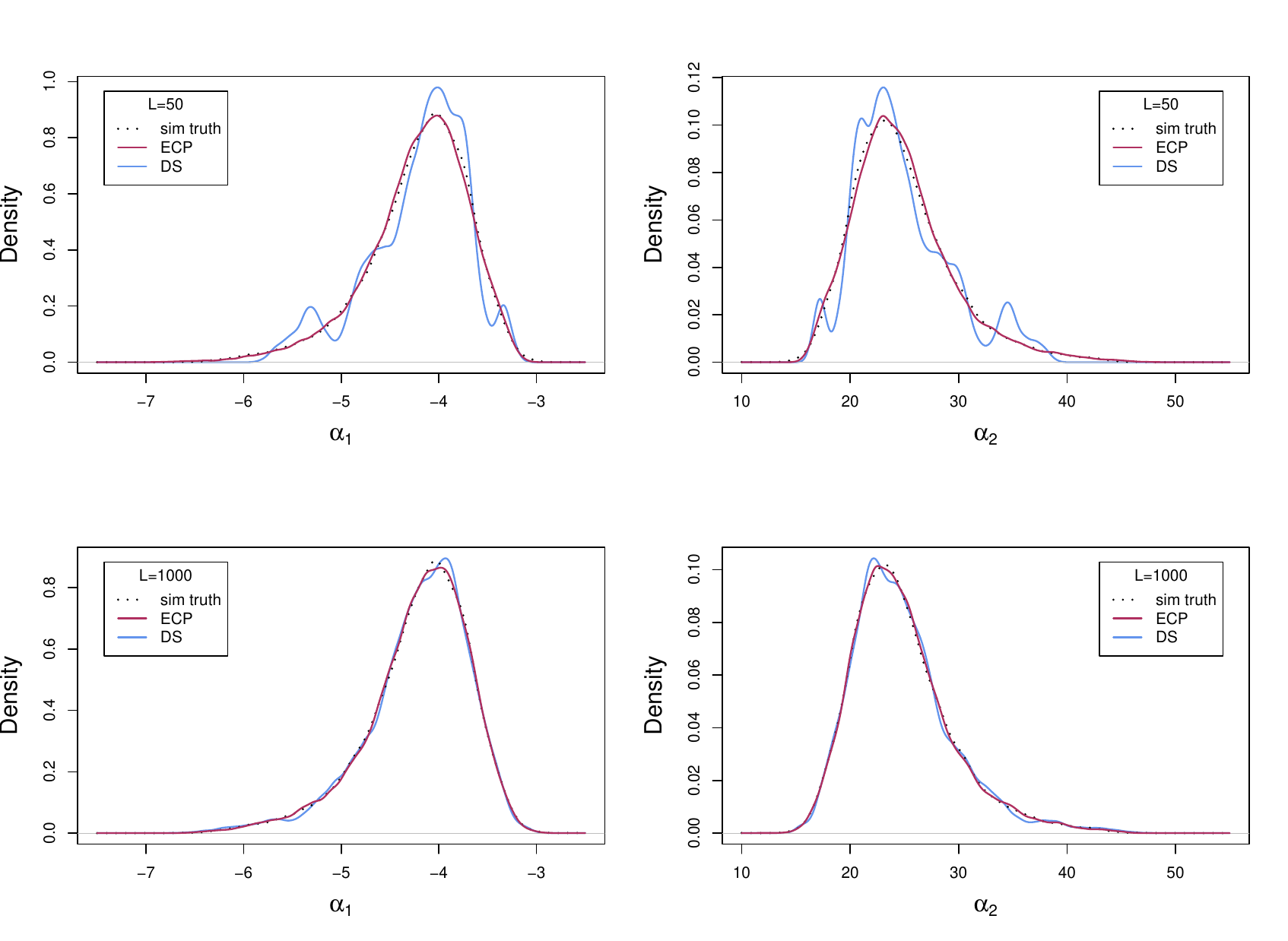}
    \caption{Estimated marginal cut-posterior distributions for $\bm \alpha$ compared to simulated ground truth. At $L=50$ the DS approximation to the cut-posterior is poor, with multimodality due to the small $L$ and relatively large $m$. The posterior approximated by ECP is much more accurate. At $L=1000$ the DS approximation is significantly improved, but still less accurate than ECP.}
    \label{fig:plummer_post}
\end{figure}

\Cref{fig:plummer_KS} shows the KS distance as compared to the simulated ground truth for DS, DS with a normal approximation, and ECP. The ECP approach can provide a much better approximation to $\pi_\text{cut}(\bm{\alpha}|\bm{Y})$ especially for small budgets. At $L=50$ the median performance of ECP is significantly better than that of DS. Only as the budget becomes large does the performance of DS approach that of ECP. At $L=1000$ the median performance of ECP and DS are fairly similar. In \cref{fig:plummer_post} the marginal cut-posterior density estimates are shown for $L=50$ in the first row and $L=1000$ in the second row. The DS approach clearly suffers at low budgets while the ECP approach provides an accurate approximation even at $L=50$. DS with a normal approximation is preferable to DS only for the smallest budget, but generally performs quite poorly and is not improved by increasing the budget as the true cut-posterior is far from Gaussian. A comparison of the bivariate densities is given in section 2 of the supplementary material. Note that this example is augmented in a way that we do not expect to recover the posterior distributions from \cite{plummer2015}.

\section{Practical Recommendations for Sampling from $\pi_\text{cut}$}\label{sec:practical}


In this section we give some practical guidance for sampling from the cut-distribution. In doing so we make it more clear which problems ECP is designed for, and for which DS may be the preferred approach.

A motivating assumption for ECP is that sampling from the prior distribution is fast, but sampling from the conditional posterior distribution is slow. This has been discussed at length and the computational complexity of both ECP and DS is dominated by $L$. If this assumption is violated, it may be that choice of $m$ and $M$ in the ECP algorithm are also relevant to computation time. Recall that $m$ is the number of samples from the conditional posterior distribution which can be acquired for each $\bm \gamma$ in phase 1 of \cref{alg:ECP}. These $m$ samples are used to estimate the parameters for $\mathcal F$. $M$ is the number of samples of $\bm \gamma$ which are used for prediction in phase 2 of the ECP algorithm, it is also the number of mixture components in the final approximation of the cut-distribution. 

\Cref{tbl:cut_sample_choices} gives insight into the choice of sampling algorithms and selection of $L,m,M$ in 4 distinct sub-cases related to the efficiency of sampling the prior and conditional distributions. Importantly, if sampling from the conditional posterior is fast, the ECP algorithm is probably not necessary and DS is sufficient.

\begin{table}[h!]
\centering
\renewcommand{\arraystretch}{2} 
\setlength{\tabcolsep}{12pt}    

\begin{tabular}{c|c|c|c|}
  \multicolumn{2}{c}{} & \multicolumn{2}{c}{$\mathbf{p(\bm\alpha | \bm\gamma, \bm y)}$} \\ \cline{3-4}
  \multicolumn{2}{c|}{} & \textbf{Inexpensive} & \textbf{Expensive} \\ \cline{2-4}
\multirow{2}{*}{$\mathbf{p(\bm\gamma | \bm z)}$} & \textbf{Inexpensive} & \parbox[c]{3.2cm}
{\centering  {\scriptsize {\color{white}.} \\ {\bf DS with DOE:} \it  When both sampling phases are fast, there is no need for the ECP algorithm. Simply apply DS with a massive $L$ and $m=1$. A larger $m$ could be used due to the large $L$, however there is no need. The use of DOE for $\bm \gamma$ is still recommended to improve the stability and accuracy of DS.\\[2ex]}} & \parbox[c]{3.55cm}
{\centering  {\scriptsize {\color{white}.} \\ {\bf ECP:} \it  A large sample set of $\bm \gamma$ is readily available through random sampling or MCMC which can be thinned to a small $L$ number of samples using support points. The large sample set can be used as the $M$ samples in phase 2. $m$ is likely to be relatively small, and so bootstrapping can be used to improve parameter estimates. $m$ can be greater than 1 because $M$ is large.\\[2ex]}} \\ \cline{2-4}
 & \textbf{Expensive} & \parbox[c]{3.2cm}
{\centering  {\scriptsize {\color{white}.} \\ {\bf DS with MinED:} \it  If sampling $\bm \gamma$ is expensive enough that a large MCMC is not feasible, SP and LHS sampling are not likely to be available. In this case, MinED can provide the $L$ samples and set $m \ll L$. $L$ should be set as large as computationally feasible. Since sampling $\bm \alpha$ is Inexpensive, consider using a kernel density estimate or parametric fit to the MinED sample set to generate a large number of samples (increasing $L$). 
\\[2ex]}} & \parbox[c]{3.2cm}
{\centering  {\scriptsize {\color{white}.} \\ {\bf ECP with MinED:} \it  Begin with a small budget $L$ and gather the initial samples of $\bm \gamma$ using MinED. MCMC can then be used to gather $m$ samples of $\bm \alpha$. Depending on the added expense of running longer chains, $m$ may need to be fairly small. A faster approach could be to again use MinED to gather these $m$ samples. 
To get a large $M$ number of samples of $\bm \gamma$, use a kernel density estimate or parametric fit from the MinED samples in the first phase.\\[2ex]}} \\ \cline{2-4}
\end{tabular}
\caption{Strategies for sampling from the cut-posterior depending on the computational requirements of sampling from the prior and the conditional posterior distribution.}
\label{tbl:cut_sample_choices}
\end{table}

The computational complexity of the ECP algorithm when $m$ and $M$ are considered is $\mathcal{O}(\frac{L+M}{P}M_{\gamma} + \frac{m}{P}M_{\alpha} + \frac{r}{P}L^3)$. Here $M_{\gamma}$ is the cost of an MCMC chain targeting $\bm \gamma$, yielding $\frac{L+M}{P}$ samples per chain. $L+M$ samples of $\bm \gamma$ are needed for the ECP algorithm, $L$ for phase 1 and $M$ for phase 2. Similarly, $M_{\alpha}$ is the cost of an MCMC chain targeting $\bm \alpha$ yielding $\frac{m}{P}$ samples.

\section{Discussion}\label{sec:discussion}

This work presents a new algorithm for approximating the cut-distribution and discusses important practical aspects associated with modular statistical learning. The gold standard approach for modular statistical learning is the DS algorithm where uncertainty in the cut parameters is propagated forward to uncertainty in the parameters of interest via a set of representative samples (see \cref{sec:cut_review}). We show that for the DS algorithm, if the computational budget is limited, the set of cut parameter samples ($\bm\gamma$) that are propagated forward must be chosen with care, otherwise the approximation of the cut-posterior may be poor. We find that Support Points \citep{mak2018support} is preferable to random sampling and minimax LHS for a number of applications.

The proposed ECP algorithm makes better use of available computational resources by learning an infinite mixture representation of the cut-distribution. 
We demonstrate that the proposed algorithm can approximate the cut-posterior extremely well for a number of test problems with a very small computational budget, making it an attractive alternative to the traditional DS approach when computation is limited. We also show that the ECP algorithm is far less sensitive than DS to the sampling algorithm used for $\pi(\bm\gamma|\bm z)$.

We have found the ECP algorithm to be effective whenever $\mathcal F$ is reasonably specified, and parameters $\psi_1,\ldots,\psi_r$ are estimated well. \Cref{alg:ECP} is written with MCMC for the conditional posterior distribution in mind, with the parameters $\psi_1,\ldots,\psi_r$ determined using sample based estimators. Importantly, this need not be the case. We also explored the use of the Laplace approximation for $\mathcal F$, which can provide a significant computational savings when MCMC targeting the conditional distribution is expensive. Results for the ecological example in \cref{sec:examples} using ECP with the Laplace approximation are given in section 2 of the supplementary materials. We did not explore the Laplace approximation for more difficult problems and suspect that it may struggle compared to MCMC when the conditional posterior is multimodal, or of especially high dimension due to its dependence on optimization routines.


In this work we consider only unimodal distributions for $\mathcal F$. A nontrival extension of this work could consider mixture distributions for $\mathcal F$ in order to approximate multimodal conditional posterior distributions $\pi(\bm \alpha | \bm \gamma, \bm y)$. An example of this is a Gaussian mixture for $\mathcal F$ fit using the EM algorithm. We intend to explore mixture distributions in future work.

A natural extension of this idea, is to select each $\bm\gamma^\ell$ sequentially (or block-sequentially, so that parallel resources can be used), placing it in a region of $\Gamma$ that yields the most {\it expected improvement}. This idea has been explored for many related problems involving Gaussian processes, such as in \cite{ranjan2008, roy2008, pratola2013, gramacy2011b}. In section 5 of the supplemental materials, we provide details on a sequential extension of the ECP algorithm, and we give a closed form acquisition function for the normal, Weibull and multivariate normal distributions. In our experience, the sequential procedure is not a significant improvement over the standard ECP algorithm. For a particular challenging problem, where (i) a very small budget $L$ is allocated or (ii) the MCMC sampler takes a very long time, the sequential procedure may be worthwhile. 




We propose the ECP algorithm not as a universal replacement for DS, but as a strategic alternative when computing resources are limited. While ECP may not result in significant gains in computation or accuracy for every problem, it consistently matches or exceeds the performance of DS. For instance, the DB problem in \cref{sec:DB_analysis} benefits substantially more from ECP than the modified ecological example in \cref{sec:examples}, demonstrating that different inference problems demand tailored approaches. However, we have yet to encounter a case where ECP fails to provide equal or better accuracy compared to DS. While future theoretical work could be done to formalize this observation, the evidence from our examples is quite convincing.

\section*{Acknowledgements}

This paper describes objective technical results and analysis. Any subjective views or opinions that might be expressed in the paper do not necessarily represent the views of the U.S. Department of Energy or the United States Government. 

\noindent This work was supported by a Los Alamos National Laboratories Laboratory Directed Research and Development (LDRD) grant. Further support was given by a Sandia National Laboratories
Laboratory Directed Research and Development (LDRD) grant and its ASC-ML program. This work was also supported by Simon Fraser University through Derek Bingham's NSREC Discovery grant. 





\bibliography{biblio}

\begin{thebibliography}{38}
\expandafter\ifx\csname natexlab\endcsname\relax\def\natexlab#1{#1}\fi
\providecommand{\url}[1]{\texttt{#1}}
\providecommand{\href}[2]{#2}
\providecommand{\path}[1]{#1}
\providecommand{\DOIprefix}{doi:}
\providecommand{\ArXivprefix}{arXiv:}
\providecommand{\URLprefix}{URL: }
\providecommand{\Pubmedprefix}{pmid:}
\providecommand{\doi}[1]{\href{http://dx.doi.org/#1}{\path{#1}}}
\providecommand{\Pubmed}[1]{\href{pmid:#1}{\path{#1}}}
\providecommand{\bibinfo}[2]{#2}
\ifx\xfnm\relax \def\xfnm[#1]{\unskip,\space#1}\fi
\bibitem[{Arendt et~al.(2012)Arendt, Apley and Chen}]{arendt2012}
\bibinfo{author}{Arendt, P.D.}, \bibinfo{author}{Apley, D.W.},
  \bibinfo{author}{Chen, W.}, \bibinfo{year}{2012}.
\newblock \bibinfo{title}{Quantification of model uncertainty: Calibration,
  model discrepancy, and identifiability}.
\newblock \bibinfo{journal}{Journal of Mechanical Design}
  \bibinfo{volume}{134}, \bibinfo{pages}{100908}.
\newblock \DOIprefix\doi{10.1115/1.4007390}.
\bibitem[{Arendt et~al.(2016)Arendt, Apley and Chen}]{arendt2016}
\bibinfo{author}{Arendt, P.D.}, \bibinfo{author}{Apley, D.W.},
  \bibinfo{author}{Chen, W.}, \bibinfo{year}{2016}.
\newblock \bibinfo{title}{A preposterior analysis to predict identifiability in
  the experimental calibration of computer models}.
\newblock \bibinfo{journal}{IIE Transactions} \bibinfo{volume}{48},
  \bibinfo{pages}{75--88}.
\newblock \DOIprefix\doi{10.1080/0740817X.2015.1064554}.
\bibitem[{Baldé et~al.(2023)Baldé, Damblin, Marrel, Bouloré and
  Giraldi}]{GPalphagamma2023}
\bibinfo{author}{Baldé, O.}, \bibinfo{author}{Damblin, G.},
  \bibinfo{author}{Marrel, A.}, \bibinfo{author}{Bouloré, A.},
  \bibinfo{author}{Giraldi, L.}, \bibinfo{year}{2023}.
\newblock \bibinfo{title}{Nonparametric bayesian approach for quantifying the
  conditional uncertainty of input parameters in chained numerical models}.
\newblock \URLprefix \url{https://arxiv.org/abs/2307.01111},
  \href{http://arxiv.org/abs/2307.01111}{{\tt arXiv:2307.01111}}.
\bibitem[{Beven and Freer(2001)}]{beven2001}
\bibinfo{author}{Beven, K.}, \bibinfo{author}{Freer, J.}, \bibinfo{year}{2001}.
\newblock \bibinfo{title}{Equifinality, data assimilation, and uncertainty
  estimation in mechanistic modelling of complex environmental systems using
  the glue methodology}.
\newblock \bibinfo{journal}{Journal of hydrology} \bibinfo{volume}{249},
  \bibinfo{pages}{11--29}.
\bibitem[{Binois and Gramacy(2021)}]{binois2021hetgp}
\bibinfo{author}{Binois, M.}, \bibinfo{author}{Gramacy, R.B.},
  \bibinfo{year}{2021}.
\newblock \bibinfo{title}{hetgp: Heteroskedastic gaussian process modeling and
  sequential design in r}.
\newblock \bibinfo{journal}{Journal of Statistical Software}
  \bibinfo{volume}{98}, \bibinfo{pages}{1--44}.
\bibitem[{Brown and Hund(2018)}]{brown2018}
\bibinfo{author}{Brown, J.L.}, \bibinfo{author}{Hund, L.B.},
  \bibinfo{year}{2018}.
\newblock \bibinfo{title}{Estimating material properties under extreme
  conditions by using bayesian model calibration with functional outputs}.
\newblock \bibinfo{journal}{Journal of the Royal Statistical Society: Series C
  (Applied Statistics)} \bibinfo{volume}{67}, \bibinfo{pages}{1023--1045}.
\newblock \DOIprefix\doi{10.1111/rssc.12273}.
\bibitem[{Carnell and Carnell(2016)}]{carnell2016package}
\bibinfo{author}{Carnell, R.}, \bibinfo{author}{Carnell, M.R.},
  \bibinfo{year}{2016}.
\newblock \bibinfo{title}{Package ‘lhs’}.
\newblock \bibinfo{journal}{CRAN. https://cran. rproject.
  org/web/packages/lhs/lhs. pdf} .
\bibitem[{Efron(1982)}]{efron1982}
\bibinfo{author}{Efron, B.}, \bibinfo{year}{1982}.
\newblock \bibinfo{title}{The jackknife, the bootstrap, and other resampling
  plans}. volume~\bibinfo{volume}{38}.
\newblock \bibinfo{publisher}{Siam}.
\bibitem[{Fiszeder and Orzeszko(2021)}]{covregression}
\bibinfo{author}{Fiszeder, P.}, \bibinfo{author}{Orzeszko, W.},
  \bibinfo{year}{2021}.
\newblock \bibinfo{title}{Covariance matrix forecasting using support vector
  regression}.
\newblock \bibinfo{journal}{Applied intelligence} \bibinfo{volume}{51},
  \bibinfo{pages}{7029--7042}.
\bibitem[{Francom(2020)}]{BASS_package}
\bibinfo{author}{Francom, D.}, \bibinfo{year}{2020}.
\newblock \bibinfo{title}{BASS: Bayesian Adaptive Spline Surfaces}.
\newblock \URLprefix \url{https://CRAN.R-project.org/package=BASS}.
  \bibinfo{note}{r package version 1.2.0}.
\bibitem[{Freedman(1963)}]{Freedman1963}
\bibinfo{author}{Freedman, D.A.}, \bibinfo{year}{1963}.
\newblock \bibinfo{title}{On the asymptotic behavior of bayes' estimates in the
  discrete case}.
\newblock \bibinfo{journal}{The Annals of Mathematical Statistics} ,
  \bibinfo{pages}{1386--1403}.
\bibitem[{Friedman et~al.(2008)Friedman, Hastie and Tibshirani}]{friedman2008}
\bibinfo{author}{Friedman, J.}, \bibinfo{author}{Hastie, T.},
  \bibinfo{author}{Tibshirani, R.}, \bibinfo{year}{2008}.
\newblock \bibinfo{title}{Sparse inverse covariance estimation with the
  graphical lasso}.
\newblock \bibinfo{journal}{Biostatistics} \bibinfo{volume}{9},
  \bibinfo{pages}{432--441}.
\bibitem[{Goodfellow et~al.(2016)Goodfellow, Bengio and
  Courville}]{goodfellow2016deep}
\bibinfo{author}{Goodfellow, I.}, \bibinfo{author}{Bengio, Y.},
  \bibinfo{author}{Courville, A.}, \bibinfo{year}{2016}.
\newblock \bibinfo{title}{Deep learning}.
\newblock \bibinfo{publisher}{MIT press}.
\bibitem[{Gramacy and Polson(2011)}]{gramacy2011b}
\bibinfo{author}{Gramacy, R.B.}, \bibinfo{author}{Polson, N.G.},
  \bibinfo{year}{2011}.
\newblock \bibinfo{title}{Particle learning of gaussian process models for
  sequential design and optimization}.
\newblock \bibinfo{journal}{Journal of Computational and Graphical Statistics}
  \bibinfo{volume}{20}, \bibinfo{pages}{102--118}.
\bibitem[{Helton and Davis(2003)}]{helton2003latin}
\bibinfo{author}{Helton, J.C.}, \bibinfo{author}{Davis, F.J.},
  \bibinfo{year}{2003}.
\newblock \bibinfo{title}{Latin hypercube sampling and the propagation of
  uncertainty in analyses of complex systems}.
\newblock \bibinfo{journal}{Reliability Engineering \& System Safety}
  \bibinfo{volume}{81}, \bibinfo{pages}{23--69}.
\bibitem[{Higham(1988)}]{HIGHAM1988103}
\bibinfo{author}{Higham, N.J.}, \bibinfo{year}{1988}.
\newblock \bibinfo{title}{Computing a nearest symmetric positive semidefinite
  matrix}.
\newblock \bibinfo{journal}{Linear Algebra and its Applications}
  \bibinfo{volume}{103}, \bibinfo{pages}{103--118}.
\newblock \DOIprefix\doi{https://doi.org/10.1016/0024-3795(88)90223-6}.
\bibitem[{Jacob et~al.(2017)Jacob, Murray, Holmes and
  Robert}]{jacob2017BetterTogether}
\bibinfo{author}{Jacob, P.E.}, \bibinfo{author}{Murray, L.M.},
  \bibinfo{author}{Holmes, C.C.}, \bibinfo{author}{Robert, C.P.},
  \bibinfo{year}{2017}.
\newblock \bibinfo{title}{Better together? statistical learning in models made
  of modules}.
\newblock \bibinfo{journal}{arXiv preprint arXiv:1708.08719} .
\bibitem[{Jacob et~al.(2020)Jacob, O'Leary and
  Atchad{\'e}}]{jacob2020UnbiasedMarkov}
\bibinfo{author}{Jacob, P.E.}, \bibinfo{author}{O'Leary, J.},
  \bibinfo{author}{Atchad{\'e}, Y.F.}, \bibinfo{year}{2020}.
\newblock \bibinfo{title}{Unbiased markov chain monte carlo methods with
  couplings}.
\newblock \bibinfo{journal}{Journal of the Royal Statistical Society: Series B
  (Statistical Methodology)} \bibinfo{volume}{82}, \bibinfo{pages}{543--600}.
\bibitem[{Jean et~al.(2009)Jean, Noguere, Habert and Iooss}]{jean2009}
\bibinfo{author}{Jean, C.D.S.}, \bibinfo{author}{Noguere, G.},
  \bibinfo{author}{Habert, B.}, \bibinfo{author}{Iooss, B.},
  \bibinfo{year}{2009}.
\newblock \bibinfo{title}{A monte carlo approach to nuclear model parameter
  uncertainties propagation}.
\newblock \bibinfo{journal}{Nuclear science and engineering}
  \bibinfo{volume}{161}, \bibinfo{pages}{363--370}.
\bibitem[{Joseph et~al.(2015)Joseph, Dasgupta, Tuo and Wu}]{joseph2015mined}
\bibinfo{author}{Joseph, V.R.}, \bibinfo{author}{Dasgupta, T.},
  \bibinfo{author}{Tuo, R.}, \bibinfo{author}{Wu, C.J.}, \bibinfo{year}{2015}.
\newblock \bibinfo{title}{Sequential exploration of complex surfaces using
  minimum energy designs}.
\newblock \bibinfo{journal}{Technometrics} \bibinfo{volume}{57},
  \bibinfo{pages}{64--74}.
\bibitem[{Katzfuss et~al.(2022)Katzfuss, Guinness and Lawrence}]{sVecchia}
\bibinfo{author}{Katzfuss, M.}, \bibinfo{author}{Guinness, J.},
  \bibinfo{author}{Lawrence, E.}, \bibinfo{year}{2022}.
\newblock \bibinfo{title}{Scaled vecchia approximation for fast computer-model
  emulation}.
\newblock \bibinfo{journal}{SIAM/ASA Journal on Uncertainty Quantification}
  \bibinfo{volume}{10}, \bibinfo{pages}{537--554}.
\bibitem[{Little(1992)}]{little1992}
\bibinfo{author}{Little, R.J.}, \bibinfo{year}{1992}.
\newblock \bibinfo{title}{Regression with missing x's: a review}.
\newblock \bibinfo{journal}{Journal of the American statistical association}
  \bibinfo{volume}{87}, \bibinfo{pages}{1227--1237}.
\bibitem[{Liu et~al.(2009)Liu, Bayarri and Berger}]{liu2009}
\bibinfo{author}{Liu, F.}, \bibinfo{author}{Bayarri, M.J.},
  \bibinfo{author}{Berger, J.O.}, \bibinfo{year}{2009}.
\newblock \bibinfo{title}{Modularization in bayesian analysis, with emphasis on
  analysis of computer models}.
\newblock \bibinfo{journal}{Bayesian Analysis} \bibinfo{volume}{4},
  \bibinfo{pages}{119--150}.
\bibitem[{Liu et~al.(2020)Liu, Ong, Shen and Cai}]{liu2020gaussian}
\bibinfo{author}{Liu, H.}, \bibinfo{author}{Ong, Y.S.}, \bibinfo{author}{Shen,
  X.}, \bibinfo{author}{Cai, J.}, \bibinfo{year}{2020}.
\newblock \bibinfo{title}{When gaussian process meets big data: A review of
  scalable gps}.
\newblock \bibinfo{journal}{IEEE Transactions on Neural Networks and Learning
  Systems} \bibinfo{volume}{31}, \bibinfo{pages}{4405--4423}.
\newblock \DOIprefix\doi{10.1109/TNNLS.2019.2957109}.
\bibitem[{Liu and Goudie(2022)}]{liu2022stochastic}
\bibinfo{author}{Liu, Y.}, \bibinfo{author}{Goudie, R.J.},
  \bibinfo{year}{2022}.
\newblock \bibinfo{title}{Stochastic approximation cut algorithm for inference
  in modularized bayesian models}.
\newblock \bibinfo{journal}{Statistics and computing} \bibinfo{volume}{32},
  \bibinfo{pages}{7}.
\bibitem[{Mak and Joseph(2018)}]{mak2018support}
\bibinfo{author}{Mak, S.}, \bibinfo{author}{Joseph, V.R.},
  \bibinfo{year}{2018}.
\newblock \bibinfo{title}{{Support points}}.
\newblock \bibinfo{journal}{The Annals of Statistics} \bibinfo{volume}{46},
  \bibinfo{pages}{2562 -- 2592}.
\newblock \DOIprefix\doi{10.1214/17-AOS1629}.
\bibitem[{McKay(1992)}]{mckay1992latin}
\bibinfo{author}{McKay, M.D.}, \bibinfo{year}{1992}.
\newblock \bibinfo{title}{Latin hypercube sampling as a tool in uncertainty
  analysis of computer models}, in: \bibinfo{booktitle}{Proceedings of the 24th
  conference on Winter simulation}, pp. \bibinfo{pages}{557--564}.
\bibitem[{Plummer(2015)}]{plummer2015}
\bibinfo{author}{Plummer, M.}, \bibinfo{year}{2015}.
\newblock \bibinfo{title}{Cuts in bayesian graphical models}.
\newblock \bibinfo{journal}{Statistics and Computing} \bibinfo{volume}{25},
  \bibinfo{pages}{37--43}.
\bibitem[{Pratola et~al.(2013)Pratola, Sain, Bingham, Wiltberger and
  Rigler}]{pratola2013}
\bibinfo{author}{Pratola, M.T.}, \bibinfo{author}{Sain, S.R.},
  \bibinfo{author}{Bingham, D.}, \bibinfo{author}{Wiltberger, M.},
  \bibinfo{author}{Rigler, E.J.}, \bibinfo{year}{2013}.
\newblock \bibinfo{title}{Fast sequential computer model calibration of large
  nonstationary spatial-temporal processes}.
\newblock \bibinfo{journal}{Technometrics} \bibinfo{volume}{55},
  \bibinfo{pages}{232--242}.
\bibitem[{Ranjan et~al.(2008)Ranjan, Bingham and Michailidis}]{ranjan2008}
\bibinfo{author}{Ranjan, P.}, \bibinfo{author}{Bingham, D.},
  \bibinfo{author}{Michailidis, G.}, \bibinfo{year}{2008}.
\newblock \bibinfo{title}{Sequential experiment design for contour estimation
  from complex computer codes}.
\newblock \bibinfo{journal}{Technometrics} \bibinfo{volume}{50},
  \bibinfo{pages}{527--541}.
\bibitem[{Roy(2008)}]{roy2008}
\bibinfo{author}{Roy, S.}, \bibinfo{year}{2008}.
\newblock \bibinfo{title}{Sequential-Adaptive Design of Computer Experiments
  for the Estimation of Percentiles}.
\newblock Ph.D. thesis. The Ohio State University.
\bibitem[{Rumsey et~al.(2020)Rumsey, Huerta, Brown and
  Hund}]{rumsey2020dealing}
\bibinfo{author}{Rumsey, K.}, \bibinfo{author}{Huerta, G.},
  \bibinfo{author}{Brown, J.}, \bibinfo{author}{Hund, L.},
  \bibinfo{year}{2020}.
\newblock \bibinfo{title}{Dealing with measurement uncertainties as nuisance
  parameters in bayesian model calibration}.
\newblock \bibinfo{journal}{SIAM/ASA Journal on Uncertainty Quantification}
  \bibinfo{volume}{8}, \bibinfo{pages}{1287--1309}.
\bibitem[{Rumsey(2020)}]{rumsey2020methods}
\bibinfo{author}{Rumsey, K.N.}, \bibinfo{year}{2020}.
\newblock \bibinfo{title}{Methods of uncertainty quantification for physical
  parameters}.
\newblock Ph.D. thesis. The University of New Mexico.
\bibitem[{Sparapani et~al.(2021)Sparapani, Spanbauer and McCulloch}]{bart}
\bibinfo{author}{Sparapani, R.}, \bibinfo{author}{Spanbauer, C.},
  \bibinfo{author}{McCulloch, R.}, \bibinfo{year}{2021}.
\newblock \bibinfo{title}{Nonparametric machine learning and efficient
  computation with bayesian additive regression trees: The bart r package}.
\newblock \bibinfo{journal}{Journal of Statistical Software}
  \bibinfo{volume}{97}, \bibinfo{pages}{1–66}.
\newblock \DOIprefix\doi{10.18637/jss.v097.i01}.
\bibitem[{Tierney and Kadane(1986)}]{laplace_approx}
\bibinfo{author}{Tierney, L.}, \bibinfo{author}{Kadane, J.B.},
  \bibinfo{year}{1986}.
\newblock \bibinfo{title}{Accurate approximations for posterior moments and
  marginal densities}.
\newblock \bibinfo{journal}{Journal of the American Statistical Association}
  \bibinfo{volume}{81}, \bibinfo{pages}{82--86}.
\newblock \DOIprefix\doi{10.1080/01621459.1986.10478240}.
\bibitem[{Vernon and Gosling(2023)}]{vernon2023bayesian}
\bibinfo{author}{Vernon, I.}, \bibinfo{author}{Gosling, J.P.},
  \bibinfo{year}{2023}.
\newblock \bibinfo{title}{A bayesian computer model analysis of robust bayesian
  analyses}.
\newblock \bibinfo{journal}{Bayesian Analysis} \bibinfo{volume}{18},
  \bibinfo{pages}{1367--1399}.
\bibitem[{Yu et~al.(2023)Yu, Nott and Smith}]{yu2023variational}
\bibinfo{author}{Yu, X.}, \bibinfo{author}{Nott, D.J.}, \bibinfo{author}{Smith,
  M.S.}, \bibinfo{year}{2023}.
\newblock \bibinfo{title}{Variational inference for cutting feedback in
  misspecified models}.
\newblock \bibinfo{journal}{Statistical Science} \bibinfo{volume}{38},
  \bibinfo{pages}{490--509}.
\bibitem[{Zigler et~al.(2013)Zigler, Watts, Yeh, Wang, Coull and
  Dominici}]{zigler2013model}
\bibinfo{author}{Zigler, C.M.}, \bibinfo{author}{Watts, K.},
  \bibinfo{author}{Yeh, R.W.}, \bibinfo{author}{Wang, Y.},
  \bibinfo{author}{Coull, B.A.}, \bibinfo{author}{Dominici, F.},
  \bibinfo{year}{2013}.
\newblock \bibinfo{title}{Model feedback in bayesian propensity score
  estimation}.
\newblock \bibinfo{journal}{Biometrics} \bibinfo{volume}{69},
  \bibinfo{pages}{263--273}.

\end{thebibliography}


\begin{thebibliography}{12}
\expandafter\ifx\csname natexlab\endcsname\relax\def\natexlab#1{#1}\fi
\providecommand{\url}[1]{\texttt{#1}}
\providecommand{\href}[2]{#2}
\providecommand{\path}[1]{#1}
\providecommand{\DOIprefix}{doi:}
\providecommand{\ArXivprefix}{arXiv:}
\providecommand{\URLprefix}{URL: }
\providecommand{\Pubmedprefix}{pmid:}
\providecommand{\doi}[1]{\href{http://dx.doi.org/#1}{\path{#1}}}
\providecommand{\Pubmed}[1]{\href{pmid:#1}{\path{#1}}}
\providecommand{\bibinfo}[2]{#2}
\ifx\xfnm\relax \def\xfnm[#1]{\unskip,\space#1}\fi
\bibitem[{Chen and Welch(2019)}]{chen2019}
\bibinfo{author}{Chen, H.}, \bibinfo{author}{Welch, W.J.},
  \bibinfo{year}{2019}.
\newblock \bibinfo{title}{Sequential computer experimental design for
  estimating an extreme probability or quantile}.
\newblock \bibinfo{journal}{arXiv preprint arXiv:1908.05357} .
\bibitem[{Giles(2008)}]{giles2008}
\bibinfo{author}{Giles, M.B.}, \bibinfo{year}{2008}.
\newblock \bibinfo{title}{Multilevel monte carlo path simulation}.
\newblock \bibinfo{journal}{Operations research} \bibinfo{volume}{56},
  \bibinfo{pages}{607--617}.
\bibitem[{Gramacy and Polson(2011)}]{gramacy2011b}
\bibinfo{author}{Gramacy, R.B.}, \bibinfo{author}{Polson, N.G.},
  \bibinfo{year}{2011}.
\newblock \bibinfo{title}{Particle learning of gaussian process models for
  sequential design and optimization}.
\newblock \bibinfo{journal}{Journal of Computational and Graphical Statistics}
  \bibinfo{volume}{20}, \bibinfo{pages}{102--118}.
\bibitem[{Joseph et~al.(2015)Joseph, Dasgupta, Tuo and Wu}]{Joseph2015mined}
\bibinfo{author}{Joseph, V.R.}, \bibinfo{author}{Dasgupta, T.},
  \bibinfo{author}{Tuo, R.}, \bibinfo{author}{Wu, C.J.}, \bibinfo{year}{2015}.
\newblock \bibinfo{title}{Sequential exploration of complex surfaces using
  minimum energy designs}.
\newblock \bibinfo{journal}{Technometrics} \bibinfo{volume}{57},
  \bibinfo{pages}{64--74}.
\bibitem[{Liu et~al.(2009)Liu, Bayarri and Berger}]{liu2009}
\bibinfo{author}{Liu, F.}, \bibinfo{author}{Bayarri, M.J.},
  \bibinfo{author}{Berger, J.O.}, \bibinfo{year}{2009}.
\newblock \bibinfo{title}{Modularization in bayesian analysis, with emphasis on
  analysis of computer models}.
\newblock \bibinfo{journal}{Bayesian Analysis} \bibinfo{volume}{4},
  \bibinfo{pages}{119--150}.
\bibitem[{Mak and Joseph(2018)}]{mak2018support}
\bibinfo{author}{Mak, S.}, \bibinfo{author}{Joseph, V.R.},
  \bibinfo{year}{2018}.
\newblock \bibinfo{title}{{Support points}}.
\newblock \bibinfo{journal}{The Annals of Statistics} \bibinfo{volume}{46},
  \bibinfo{pages}{2562 -- 2592}.
\newblock \DOIprefix\doi{10.1214/17-AOS1629}.
\bibitem[{McKay(1992)}]{mckay1992latin}
\bibinfo{author}{McKay, M.D.}, \bibinfo{year}{1992}.
\newblock \bibinfo{title}{Latin hypercube sampling as a tool in uncertainty
  analysis of computer models}, in: \bibinfo{booktitle}{Proceedings of the 24th
  conference on Winter simulation}, pp. \bibinfo{pages}{557--564}.
\bibitem[{Myers et~al.(2016)Myers, Lawrence, Fugate, Bowen, Ticknor, Woodring,
  Wendelberger and Ahrens}]{myers2016}
\bibinfo{author}{Myers, K.}, \bibinfo{author}{Lawrence, E.},
  \bibinfo{author}{Fugate, M.}, \bibinfo{author}{Bowen, C.M.},
  \bibinfo{author}{Ticknor, L.}, \bibinfo{author}{Woodring, J.},
  \bibinfo{author}{Wendelberger, J.}, \bibinfo{author}{Ahrens, J.},
  \bibinfo{year}{2016}.
\newblock \bibinfo{title}{Partitioning a large simulation as it runs}.
\newblock \bibinfo{journal}{Technometrics} \bibinfo{volume}{58},
  \bibinfo{pages}{329--340}.
\bibitem[{Pratola et~al.(2013)Pratola, Sain, Bingham, Wiltberger and
  Rigler}]{pratola2013}
\bibinfo{author}{Pratola, M.T.}, \bibinfo{author}{Sain, S.R.},
  \bibinfo{author}{Bingham, D.}, \bibinfo{author}{Wiltberger, M.},
  \bibinfo{author}{Rigler, E.J.}, \bibinfo{year}{2013}.
\newblock \bibinfo{title}{Fast sequential computer model calibration of large
  nonstationary spatial-temporal processes}.
\newblock \bibinfo{journal}{Technometrics} \bibinfo{volume}{55},
  \bibinfo{pages}{232--242}.
\bibitem[{Ranjan et~al.(2008)Ranjan, Bingham and Michailidis}]{ranjan2008}
\bibinfo{author}{Ranjan, P.}, \bibinfo{author}{Bingham, D.},
  \bibinfo{author}{Michailidis, G.}, \bibinfo{year}{2008}.
\newblock \bibinfo{title}{Sequential experiment design for contour estimation
  from complex computer codes}.
\newblock \bibinfo{journal}{Technometrics} \bibinfo{volume}{50},
  \bibinfo{pages}{527--541}.
\bibitem[{Roy(2008)}]{roy2008}
\bibinfo{author}{Roy, S.}, \bibinfo{year}{2008}.
\newblock \bibinfo{title}{Sequential-Adaptive Design of Computer Experiments
  for the Estimation of Percentiles}.
\newblock Ph.D. thesis. The Ohio State University.
\bibitem[{Xiong et~al.(2013)Xiong, Qian and Wu}]{xiong2013}
\bibinfo{author}{Xiong, S.}, \bibinfo{author}{Qian, P.Z.G.},
  \bibinfo{author}{Wu, C.F.J.}, \bibinfo{year}{2013}.
\newblock \bibinfo{title}{Sequential design and analysis of high-accuracy and
  low-accuracy computer codes}.
\newblock \bibinfo{journal}{Technometrics} \bibinfo{volume}{55},
  \bibinfo{pages}{37--46}.
\newblock \DOIprefix\doi{10.1080/00401706.2012.723572}.

\end{thebibliography}




\end{document}






\title{Supplementary Materials: Enhancing Approximate Modular Bayesian Inference by Emulating the Conditional Posterior}

\author[1,2]{Grant Hutchings}
\author[1]{Kellin N. Rumsey}
\author[2]{Derek Bingham}
\author[3]{Gabriel Huerta}

\address[1]{Statistical Sciences, Los Alamos National Laboratory, NM, United States}
\address[2]{Department of Statistics, Simon Fraser University, BC, Canada}
\address[3]{Statistical Sciences, Sandia National Laboratory, NM, United States}
\address[3]{Department of Mathematics and Statistics, University of New Mexico, NM, United States}

\begin{abstract}
  This supplementary material gives some more detail about the derivation in the diamond in a box example, and discusses the case where some data is less trustworthy than others. Additionally, the data for the ecological example is given so that our results may be reproduced, and we discuss the use of the Laplace approximation with ECP in the context of that example. Lastly, some details of the sequential ECP algorithm are discussed.
\end{abstract}

\maketitle



\section{Proofs}

\noindent\textbf{Assumption 3.}
For every $\epsilon > 0$ and every compact set $\Gamma_c\subseteq\Gamma$, there exists a set $G=\{\bm\gamma^1, \bm\gamma^2, \ldots \bm\gamma^L\} \subseteq \Gamma_c$, and an emulator $\Psi_j(\cdot)$ trained on this set such that
$$\sup_{\bm\gamma \in \Gamma_c}|\Psi_j(\bm\gamma) - \psi_j(\bm\gamma)| < \epsilon. $$

\begin{lemma}
If $\Psi_j(\bm\gamma)$ is a continuous function of $\bm\gamma$ which interpolates (e.g.,  $\Psi_j(\bm\gamma^k) = \psi_j(\bm\gamma^k)$ for every $\bm\gamma^k \in G$), then Assumption 3 holds.
\end{lemma}
\begin{proof}
\textit{We begin by noting that the assumption of continuity of both $\Psi(\bm\gamma)$ and $\psi(\bm\gamma)$ implies uniform continuity of both on the compact domain $\Gamma_c$. 
Let $\epsilon_1,\epsilon_2>0$. 
By uniform continuity, $\exists\;\delta_1,\delta_2>0$ s.t.}
\begin{equation*}
\begin{aligned}
\sup_{\bm\gamma \in \Gamma_c}|\Psi_j(\bm\gamma) - \psi_j(\bm\gamma)| 
=& \sup_{\bm\gamma \in \Gamma_c}|\Psi_j(\bm\gamma) - \psi_j(\bm\gamma) + \Psi_j(\bm\gamma_0) - \Psi_j(\bm\gamma_0)| \\
=& \sup_{\bm\gamma \in \Gamma_c}|(\Psi_j(\bm\gamma) - \Psi_j(\gamma_0)) + ( \Psi_j(\bm\gamma_0) - \psi_j(\bm\gamma))|\\
\leq& \sup_{\bm\gamma \in \Gamma_c}|\Psi_j(\bm\gamma) - \Psi_j(\gamma_0)| + | \psi_j(\bm\gamma_0) - \psi_j(\bm\gamma)|\\
 <& \epsilon_1+\epsilon_2,\\
\end{aligned}
\end{equation*}
where $\bm\gamma_0$ is any member of the training set s.t. $||\bm\gamma-\bm\gamma_0||_2<min\{\delta_1,\delta_2\}$.
\end{proof}

\begin{theorem}
If assumptions 1-3 hold, $\Gamma$ is compact, and the distribution function $F_{\alpha}$ of $\mathcal F$ is continuous with respect to $\psi_1,\ldots,\psi_r$ $\forall \; \alpha$, then $\pi_{\text{cut}}^M \xrightarrow{p} \pi_{\text{cut}}$ as $m,M,L\rightarrow\infty$ if $\{\bm\gamma^1,\ldots,\bm\gamma^L\}$, and $\{\bm\gamma^1,\ldots,\bm\gamma^M\}$ are i.i.d. samples from $\pi(\bm\gamma|\bm z)$.
\end{theorem}
\begin{proof}
\textit{By lemma 1 we have that}
$$\lim_{L\rightarrow\infty}\Psi_j(\bm\gamma)=\hat{\psi}_j(\bm\gamma)\; \forall \; j,\bm\gamma \in \Gamma,$$
\textit{where $\hat{\psi}_j(\bm\gamma)$ are the estimators used to train the emulator $\Psi_j(\bm \gamma)$. Furthermore, since the estimators $\hat{\psi}_j$ are consistent,}
\begin{equation*}
\begin{aligned}
&\text{Pr}\big(\lim_{L\rightarrow\infty}\lim_{m\rightarrow\infty}\Psi_j(\bm\gamma)=\psi_j(\bm\gamma)\big)=1 \; \forall \; j,\bm\gamma \in \Gamma \\
\implies&\text{Pr}\big(\lim_{L\rightarrow\infty}\lim_{m\rightarrow\infty}F_{\alpha}(\Psi_1(\bm\gamma), \ldots,\Psi_r(\bm\gamma)) = F_{\alpha}(\psi_1(\bm\gamma), \ldots, \psi_r(\bm\gamma))\big)=1\;\forall\;\bm\gamma\in\Gamma
\end{aligned}
\end{equation*}
\textit{by the continuous mapping theorem. Since $F_{\alpha}(\psi_1, \ldots, \psi_r)$ is the true distribution function for $\pi(\bm \alpha | \bm \gamma, \bm y$), we have that with probability 1,}
\begin{equation*}
\begin{aligned}
\lim_{M,L,m \rightarrow\infty} \pi_{\text{cut}}^M(\alpha|\bm y) &= \lim_{M,L,m\rightarrow\infty} \frac{1}{M}\sum_{s=1}^M \mathcal F\left(\alpha | \Psi_1(\bm\gamma^s), \ldots \Psi_r(\bm\gamma^s)\right) \\
&= \lim_{M\rightarrow\infty}\frac{1}{M}\sum_{s=1}^M F_{\alpha}(\psi_1(\bm \gamma^s), \ldots, \psi_r(\bm \gamma^s)) \\
&= \int_{\Gamma} \pi(\alpha | \bm \gamma, \bm y)\pi(\bm \gamma | \bm z) d\gamma = \pi_{\text{cut}}(\alpha|\bm y),
\end{aligned}
\end{equation*}
\textit{by the law of large numbers.}
\end{proof}
\section{Analysis of the Diamond in a Box Problem}

\subsection{Derivation of conditional posterior distribution}

\begin{align*}
    \log \pi(\alpha \mid \gamma,\bm y) &\propto 
    -\sum_{i=1}^{n_1} \frac{1}{2\sigma^2}(y_i-\alpha)^2 
    - \sum_{i=1}^{n_2} \frac{1}{2\sigma^2}(y_{i+n_1}-(\alpha+\gamma))^2 
    - \frac{1}{2\sigma^2_\alpha}(\alpha-\mu_\alpha)^2 \\
    &\propto \frac{1}{2\sigma^2} \Bigg[ 
        \sum_{i=1}^{n_1} \left(y_i^2 - 2\alpha y_i + \alpha^2\right) 
        + \sum_{i=1}^{n_2} \left( y_{i+n_1}^2 - 2\alpha y_{i+n_1} + \alpha^2 + 2\alpha \gamma + \gamma^2 \right) \\
    &\quad + \frac{\sigma^2}{\sigma^2_\alpha} \left( \alpha^2 - 2\alpha \mu_\alpha + \mu_\alpha^2 \right) \Bigg] \\
    &\propto \frac{1}{2\sigma^2} \left[ 
        \left( n_1 + n_2 + \frac{\sigma^2}{\sigma^2_\alpha} \right) \alpha^2 
        - 2\alpha \left(\sum_{i=1}^n y_i - n_2\gamma + \frac{\mu_\alpha \sigma^2}{\sigma^2_\alpha}\right) 
    \right] \\
    &\propto \frac{n_1 + n_2 + \frac{\sigma^2}{\sigma^2_\alpha}}{2\sigma^2} \left( 
        \alpha^2 - 2\alpha \frac{\sum_{i=1}^n y_i - n_2\gamma + \frac{\mu_\alpha \sigma^2}{\sigma^2_\alpha}}{n_1 + n_2 + \frac{\sigma^2}{\sigma^2_\alpha}} 
    \right) \\
    &\propto \frac{1}{2A} \left( 
        \alpha^2 - 2\alpha \left(\frac{A}{\sigma^2}\left(\sum_{i=1}^n y_i - n_2\gamma + \frac{\mu_\alpha \sigma^2}{\sigma^2_\alpha}\right)\right) 
    \right)
\end{align*}

where $A = \frac{n_1 + n_2 + \frac{\sigma^2}{\sigma^2_\alpha}}{\sigma^2}$. Let $B = A \left( \frac{\sum_{i=1}^n y_i}{\sigma^2} + \frac{\mu_\alpha}{\sigma^2_\alpha} \right)$, and $C = \frac{-A n_2}{\sigma^2}$. Then we have:

\begin{equation}
    \pi(\alpha \mid \gamma,\bm y) \sim N(B + C\gamma, A).
\end{equation}

Although equations 5-6 (from the main manuscript) are a lot to unpack, some useful information can be obtained with careful study. For instance, both distributions reduce to the same conjugate posterior when $n_2 = 0$. For fixed $n_2$ and $n_1\rightarrow \infty$, both posteriors converge in probability to the desired parameter $\alpha$. The other extreme, where $n_1$ is fixed and $n_2\rightarrow \infty$, is worth noting. Rather than converging to a constant, the posteriors instead converges in distribution to some limiting posterior. In the full Bayes case, this limiting distribution can be written as
\begin{equation}
\label{eq:Bayes_d_lim}
\lim_{n_2\rightarrow\infty} \pi(\alpha \mid \bm y) = N\left(\alpha \ \Bigg| \ 
\frac{\sigma_\gamma^2 n_1(\bar y_1-\mu_\alpha) + \sigma^2(\bar y_2-\mu_\alpha-\mu_\gamma)}
{\sigma_\gamma^2\left(n_1 + \frac{\sigma^2}{\sigma_\alpha^2}\right) + \sigma^2}, \ 
\frac{\sigma^2_\gamma}{1 + \frac{(\sigma^2 + n_1)}{(\sigma \sigma_\gamma)^{-2}}} \right).
\end{equation}

The cut-distribution has the limiting form
\begin{equation}
\label{eq:Mod_d_lim}
\lim_{n_2\rightarrow\infty} \pi_M(\alpha | \bm y) = N\left(\alpha\ | \ \bar y_2 - \mu_\alpha - \mu_\gamma, \sigma_\gamma^2 \right).
\end{equation}
In other words, the cut-distribution makes no attempt to learn about nuisance parameters and any subsequent attempt at correction will be based on the prior for $\gamma$. 

\subsection{A Suspect Module for $y$}

As discussed in the introduction, the cut-distribution may be more trustworthy than the full Bayesian approach when the model is misspecified. In the DB problem, model misspecification can arise if the scale error, which we assume is known, is incorrectly specified. In this case, it may be better to let $\gamma$ be informed by the auxiliary data $\bm z$ only, by using the cut-distribution. 

To justify this claim, we perform a simulation study, generating data ten thousand times from equation 4 (in the main manuscript) with $n_1=10$, $n_2=90$, $\alpha = 0$ and $\sigma_\gamma = 0.5$, where $\gamma$ is sampled independently from $N(0, \sigma_\gamma)$ for each simulation. The key point of these simulations, is that the error variance of the scale is assumed to be $\sigma^2 = 1$, but the actual variance, denoted $\sigma_\star^2$, will be varied from $0.1^2$ to $6.0^2$. Using empirical coverage of $95\%$ credible intervals and MSE of the point estimate as our metrics, Figure 1 summarizes the quality of inference for the three different approaches. 

\begin{figure}[H]
\centering
\includegraphics[width=.8\linewidth]{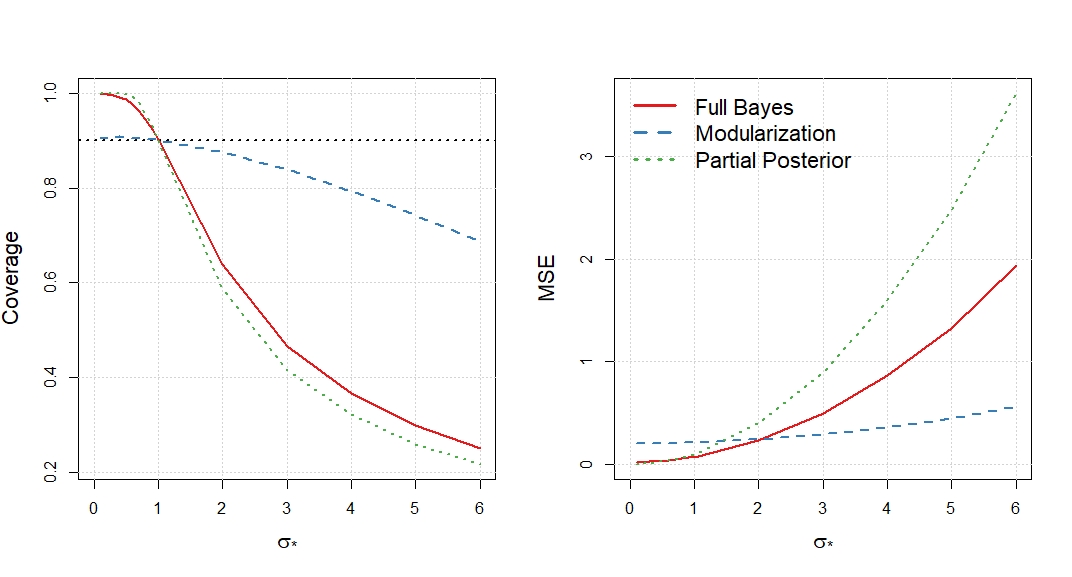}
\caption[Comparison of three approaches for misspecified prior]{Comparison of the marginal posterior (solid), the cut-distribution (dashed) and the partial posterior (dotted) for the DB problem for $\sigma_{\star} \in [0.1, 6]$. Other parameters are fixed to $n_1=10$, $n_2=90$, $\alpha=0$, $\sigma=1$, $\sigma_\gamma = 0.5$, $\sigma_\alpha = \infty$. }
\label{fig:example1}
\end{figure}

The empirical coverage of the full Bayesian and partial posterior methods drops below the nominal value as soon as $\sigma_\star > \sigma$, and they drop to less than $50\%$ empirical coverage before $\sigma_\star$ has even reached $3\sigma$. For $\sigma_\star = 3\sigma$, the cut-distribution still maintains empirical coverage of $84\%$. In terms of MSE, estimation based on the cut-distribution surpasses the other approaches for $\sigma_\star$ greater than about $2\sigma$.

\subsection{A Suspect Module for $z$}

Since the cut-distribution makes no attempt to learn about $\gamma$, it is clear that the inference will be heavily reliant on the module relating $\gamma$ and $\bm z$. If the relationship here is not well-understood, then the cut-distribution can be either worthlessly conservative (high posterior variance) or worse, completely unreliable. To demonstrate this, we repeat the simulation study from the previous section with a slight modification. This time, we assume that the variance for the scale error is correctly specified (i.e. $\sigma_\star = \sigma = 1$). Instead, we view the effect of mis-specifying the relationship between $\gamma$ and $\bm z$. This is akin to suggesting that the prior information is wrong. We can accomplish this by assuming the value $\sigma_\gamma = 0.5$ in equations 5 and 6 (main text), but generating the data (from equation 4 in main text) using $\gamma \sim N(0, \sigma_{\gamma \star})$. Figure 2 summarizes the inference for values of $\sigma_{\gamma\star} \in [0.1, 6]$, and plainly demonstrates that the cut-distribution is a poor choice for this setting. In this setting, we suggest using the partial posterior approach as discussed in \cite{liu2009}, where it is shown to have better MSE properties than the Bayesian alternative whenever $\sigma_{\gamma\star}^2 > 2\sigma_\gamma^2+\frac{1}{n_1}+\frac{1}{n_2}$. 

\begin{figure}[H]
\centering
\includegraphics[width=.8\linewidth]{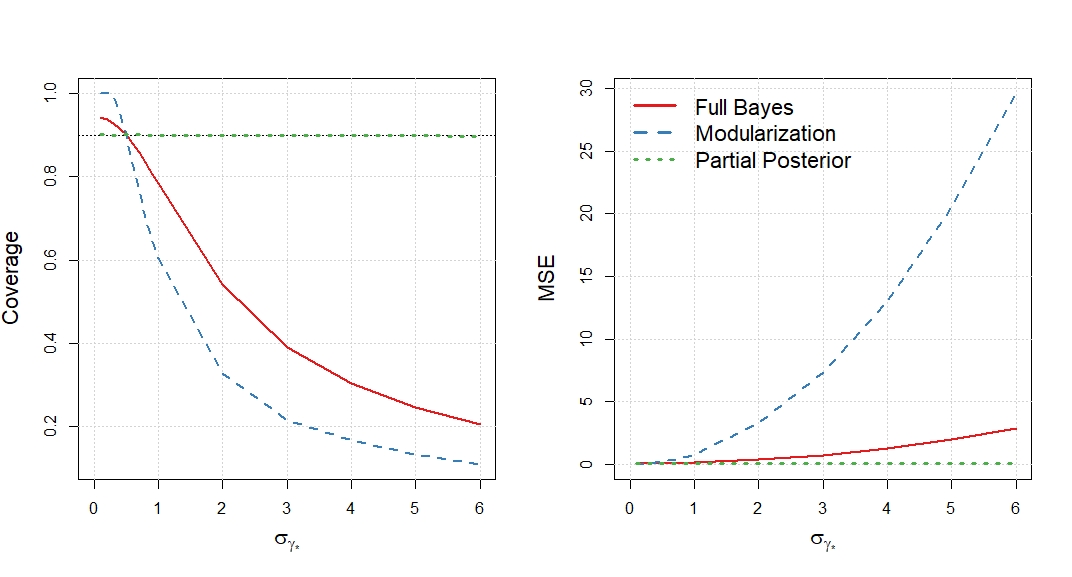}
\caption[Comparison of three approaches for misspecified prior]{Comparison of the marginal posterior (solid), the cut-distribution (dashed) and the partial posterior (dotted) for the DB problem for $\sigma_{\gamma_\star} \in [0.1, 6]$. Other parameters are fixed to $n_1=10$, $n_2=90$, $\alpha=0$, $\sigma=1$, $\sigma_\gamma = 0.5$, $\sigma_\alpha = \infty$.}
\label{fig:example2}
\end{figure}

\section{Ecological Example}

\subsection{Data and Posterior Density Estimates for cut parameters}

\begin{figure*}[h!]
    \centering
    \begin{subfigure}[t]{0.49\textwidth}  
        \centering
        \includegraphics[width=\textwidth]{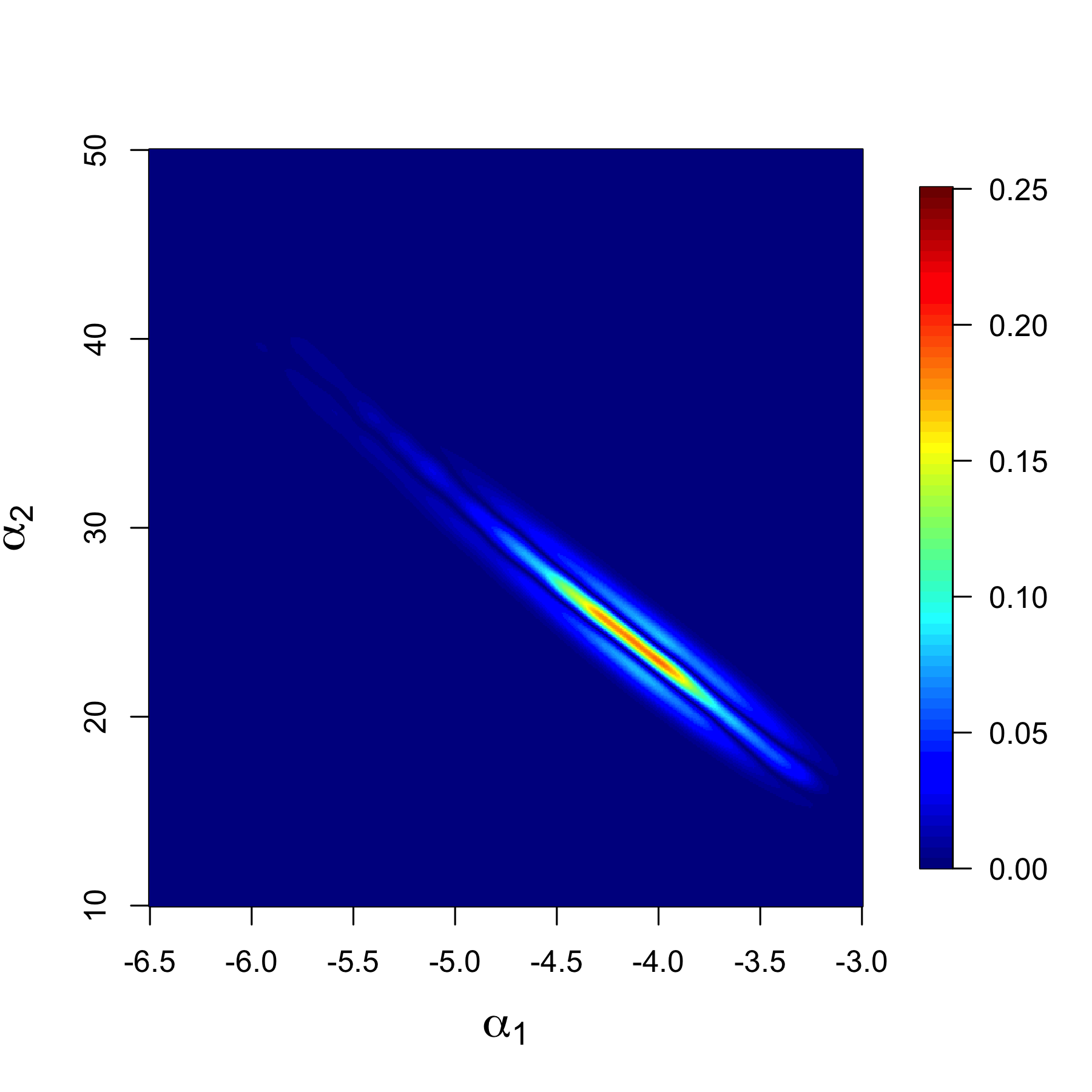}
        \caption{ECP}
        \label{fig:ECP}
    \end{subfigure}
    \begin{subfigure}[t]{0.49\textwidth}  
        \centering 
        \includegraphics[width=\textwidth]{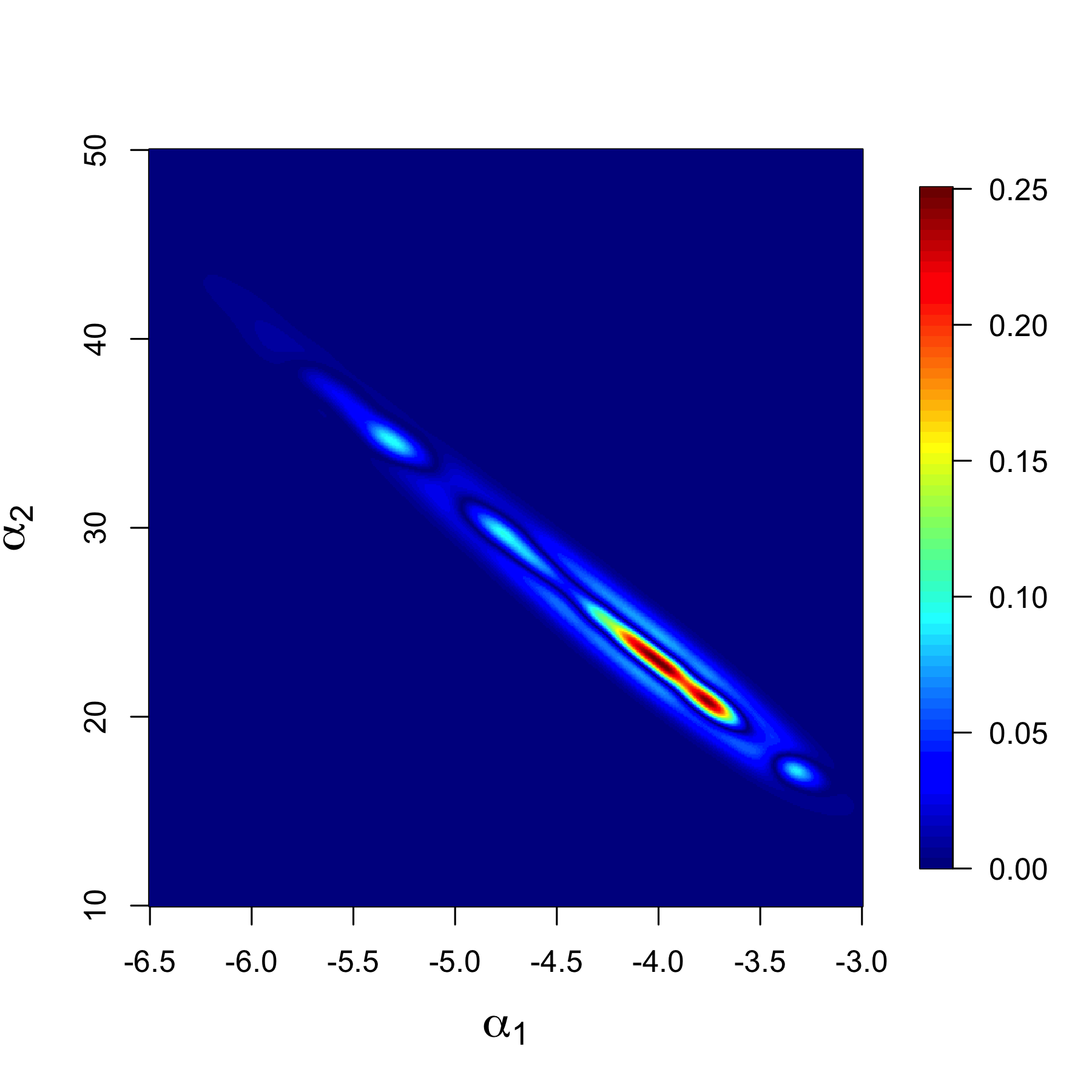}
        \caption{DS} 
        \label{fig:MI}
    \end{subfigure}
    \caption{The difference in density values between the true cut distribution and the approximated cut distributions for ECP (left) and DS (right) for $L=50$. The ECP approximation is closer to the truth and requires roughly the same amount of computation time.}
    \label{fig:2d_dens_plummer}
\end{figure*}

The data for the ecological example from Section 5.2 in the main text is given in \cref{tbl:plummer}. \Cref{fig:2d_dens_plummer} shows the difference between the true cut distribution density function and the approximated cut distributions (for ECP and DS) for the same example.

\begin{table}[!htbp]
\centering
\scalebox{0.65}{
\begin{tabular}{lccccccccccccc}
\hline
Y & 16 & 215 & 362 & 97 & 76 & 62 & 710 & 56 & 133 & 28 & 62 & 413 & 194 \\
Z & 7 & 6 & 10 & 10 & 1 & 1 & 10 & 4 & 35 & 0 & 10 & 8 & 4 \\
N & 111 & 71 & 162 & 188 & 145 & 215 & 166 & 37 & 173 & 143 & 229 & 696 & 93 \\
T & 26983 & 250930 & 829348 & 157775 & 150467 & 352445 & 553066 & 26751 & 75815 & 150302 & 354993 & 3683043 & 507218 \\
$C^1$ & 0.30& 0.43& 0.51& 0.41& 0.05& 0.55& 0.48& 0.46& 0.54& 0.40& 0.59& 0.39& 0.48 \\
$C^2$ & 0.91& 1.00& 0.90& 0.88& 0.54& 0.47& 0.91& 1.00& 1.00& 0.54& 0.81& 0.41& 0.82 \\
$C^3$ & 0.55& 0.49& 0.37& 0.58& 0.79& 0.42& 0.56& 0.56& 0.59& 0.66& 0.44& 0.46& 0.57 \\
$C^4$ & 0.58& 0.77& 0.48& 0.50& 0.40& 0.63& 0.47& 0.38& 0.25& 0.60& 0.51& 0.48& 0.49 \\
$C^5$ & 0.59& 0.39& 0.47& 0.46& 0.37& 0.35& 0.67& 0.49& 0.43& 0.41& 0.55& 0.75& 0.47 \\
\hline
\end{tabular}
}
\caption{Data used in Ecological example.}
\label{tbl:plummer}
\end{table}

\begin{figure}
    \centering
    \includegraphics[width=0.95\linewidth]{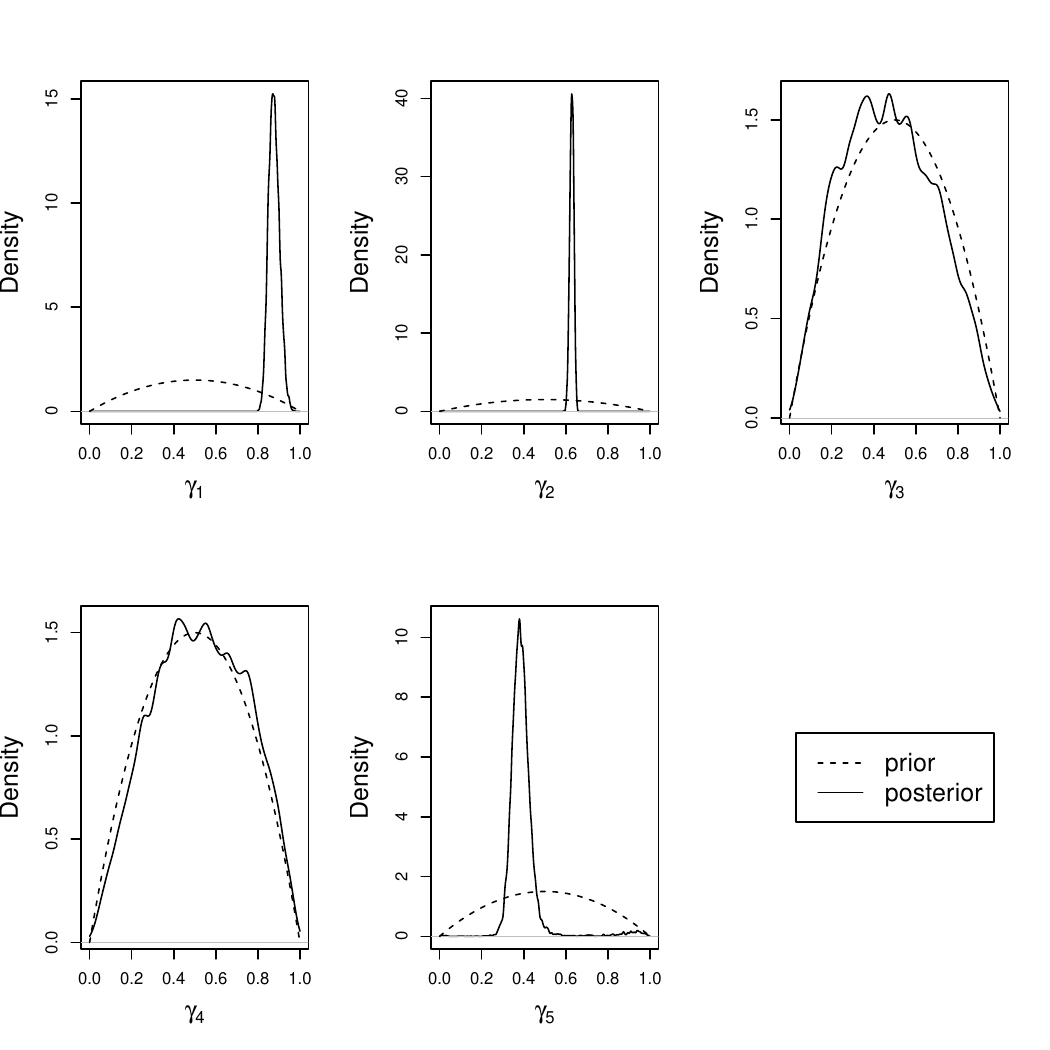}
    \caption{Prior distributions and marginal posterior density esimates for $\bm\gamma$ given $\bm z$}
    \label{fig:plummer_gamma_post}
\end{figure}

Prior distributions and marginal posterior density estimates for $\bm\gamma$ given $\bm z$ are shown in \cref{fig:plummer_gamma_post}.

\subsection{Laplace approximation}
\begin{figure}
    \centering
    \includegraphics[width=0.95\linewidth]{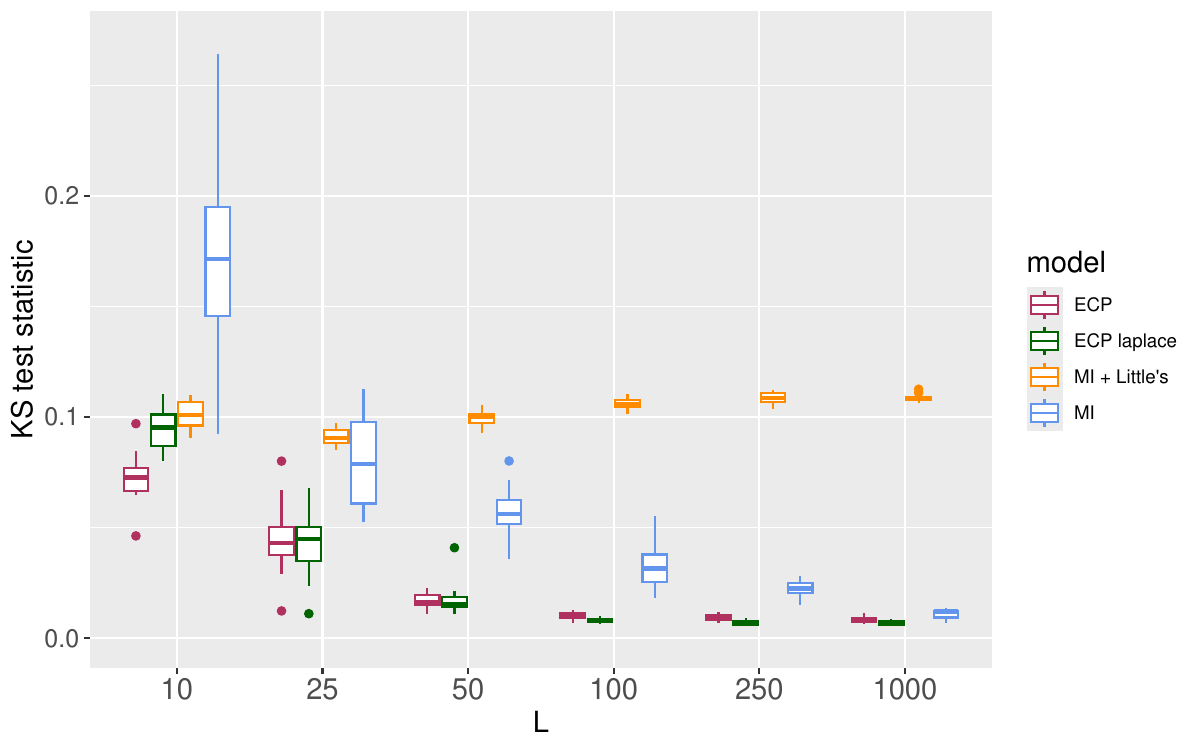}
    \caption{This figure reproduces Figure 5 in the manuscript with the addition of ECP using the Laplace approximation. The Laplace approximation seems to struggle at very low computational budgets, but may be a slightly improvement for larger computational budgets.}
    \label{fig:ecp_laplace}
\end{figure}

In the manuscript we discuss that if $\mathcal F$ is chosen to be normal, the Laplace approximation can be used to reduce computation time by avoiding MCMC targeting the conditional posterior. In \cref{fig:ecp_laplace} we add ECP with the Laplace approximation to Figure 5 from the manuscript. The difference in accuracy between ECP and ECP with the Laplace approximation is significantly different for $L=10$, but otherwise quite similar. These results indicate that the Laplace approximation is a reasonable alternative to sample based inference for a normally distributed $\mathcal F$, and may be a slight improvement for larger $L$.

\newpage
\section{Inflating the Variance Across a Bounded Space}

Consider $\bm\gamma_1, \ldots, \bm\gamma_L \sim \pi(\bm\gamma|\bm z)$ and suppose each element of $\bm\gamma$ has marginal support $\gamma_j \in [a_j, b_j]$, $j=1,\ldots, q$. Let $m_j$ and $s_j$ be the mean and standard deviation (respectively) of $\gamma_{j,1}, \ldots, \gamma_{j,L}$. When $a_j = -\infty$ and $b_j = \infty$, the variance of the training samples can be inflated by computing
$$\gamma^\star_{j,\ell} = m_j + (1+\omega)\left(\gamma_{j,\ell}-m_j\right)$$
for some variance inflation factor $\omega > 0$. When $a_j$ or $b_j$ is finite, however, this approach can lead to values of $\bm\gamma \notin \Gamma$. In this case, we propose the following ``power-likelihood" approach for variance inflation, which can reduce potential for extrapolation in the ECP algorithm. The initial sample of values is taken from the flattened ``power" distribution
$$\bm\gamma_1, \ldots, \bm\gamma_L \sim \pi(\bm\gamma|\bm z)^\omega$$
for some $\omega \in [0, 1]$. This can be accomplished numerically by (i) fitting a kernel density estimate (KDE) to a set of initial samples with $\omega = 1$, (ii) powering the density values of the KDE with $\omega \in (0, 1)$, (iii) sampling a new set of values from the altered KDE (using the empeirical CDF, for instance). This new set of samples can be used as the final set of samples or as initial samples for support points \cite{mak2018support}, minimum energy designs \cite{Joseph2015mined}, or for Latin hypercube sampling \cite{mckay1992latin}. In both cases, we recommend a choice of $\omega$ which leads to about a 5-20\% increase in the standard deviation of the samples. Note that less variance inflation is needed for larger values of $L$ and for smaller values of $q$. It is worth noting that support points and minimum energy designs can both alleviate the need for inflation, as they tend to do a reasonable job of naturally covering the tails of the distribution. 

\begin{figure}[H]
\centering
\includegraphics[width=.8\linewidth]{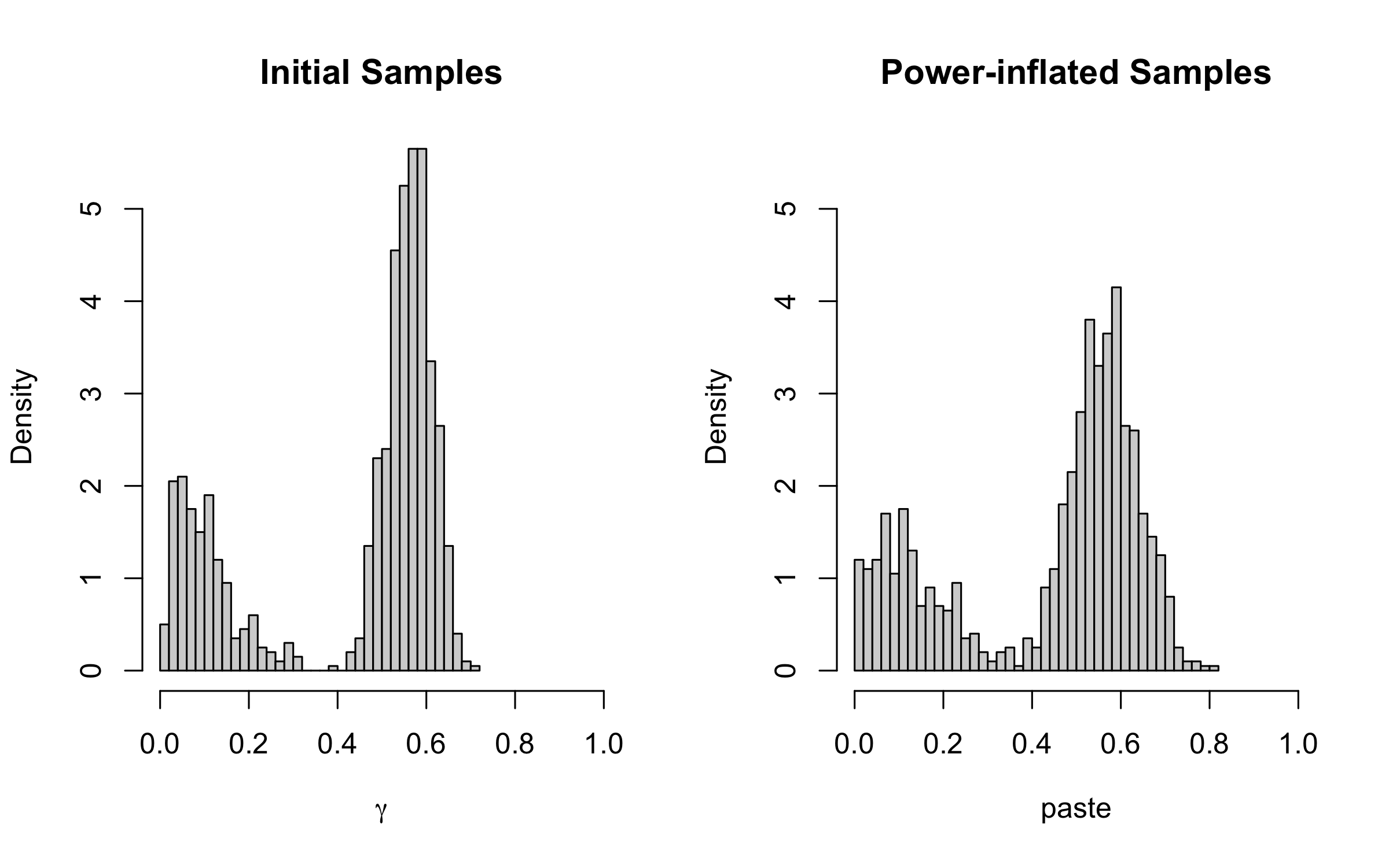}
\caption[Comparison of three approaches for misspecified prior]{Initial samples (left) and variance-inflated samples (right) for a hypothetical marginal distribution $\pi(\gamma_1|\bm z)$ with $\omega=0.75$.}
\label{fig:varinf}
\end{figure}

\Cref{fig:varinf} illustrates this approach with $\omega = 0.75$ for a hypothetical marginal distribution $\pi(\gamma_1|\bm z)$. 

\section{Sequential ECP Algorithm: Acquisition Functions}

A common idea for improving performance in the design of computer experiments, is to sequentialize tasks in order to maximize the use of precious and limited computational resources. Specific applications include (i) estimation of percentiles \cite{roy2008, chen2019}, (ii) contour estimation and optimization \cite{pratola2013, ranjan2008, gramacy2011b}, (iii) multi-fidelity modeling \cite{xiong2013, giles2008} and (iv) {\it in situ} applications \cite{myers2016}. The idea is simply that certain regions of the parameter space $\Gamma$ may be easier to learn about than others, and our budget can be more spent more effectively by choosing $\bm\gamma_\ell$ to be the location which will add the most information. This notion of equitable spending is worth discussing for ECP, although we note that the non-sequential version of our algorithm is faster (less overhead), parallelizable and sufficient for every application we have considered. In higher dimensional problems however, this extension may be valuable. 

In the first step of the ECP algorithm, the set of modularization {\it locations} ($\bm\gamma^1, \cdots $ $\bm\gamma^L$) are chosen simultaneously with respect to the prior. We want to transition to the case where $\bm\gamma^\ell$ is not chosen until the previous $\ell-1$ locations have been assessed. We begin with a {\it build phase}, in which the first three steps of ECP are executed but using $L_0$ in place of $L$ (with $L_0 < L$). 

Once the build phase is complete, the remaining budget is $L-L_0$ and the goal is to choose the remaining locations sequentially so that the Gaussian process emulators $\hat\psi_1(\cdot|\mathcal D), \cdots \hat\psi_r(\cdot|\mathcal D)$ can be learned for all reasonable values of $\bm\gamma$ as efficiently as possible. Following \cite{roy2008, pratola2013, ranjan2008}, we will define an acquisition function $I(\bm\gamma | \mathcal D)$ which is used to evaluate candidate locations, selecting the $\ell^{th}$ location as 
\begin{equation}
\label{eq:imp_gamma}
\bm\gamma_\ell = \argmax_{\bm\gamma \in \Gamma} I\left(\bm\gamma | \mathcal D_{\ell-1}\right),
\end{equation}
where $\mathcal D_\ell$ represents all of the relevant information we have acquired up to time $\ell$. Before defining the acquisition function for sequential ECP, we should discuss a few of the associated challenges. 

\begin{itemize}
\item[i)] If the location space $\Gamma$ is bounded, then it may be possible to search the entire space as in \cref{eq:imp_gamma}. When the space is unbounded finding the best location becomes more difficult. The prior distribution $\pi_\gamma(\cdot)$ needs to be accounted for in some way, to ensure that we are not wasting resources exploring locations in $\Gamma$ which are not "reasonable" with respect to the prior. One simple way to handle this problem is to replace the optimization space $\Gamma$ with a finite set of {\it candidate values} $\Gamma_c$ which contains $N_c$ random draws from the prior distribution $\pi_\gamma(\cdot)$. If $N_c$ is large enough, the chosen location can be expected to correspond to good improvement, while being constrained to a region of prior plausibility.
\item[ii)] It seems reasonable that the new modularization parameter should be placed in a location which reduces the uncertainty in the GP emulators $\hat\psi_j(\cdot)$. It may be the case that a value which maximizes the improvement for the $j^{th}$ GP has little value with respect to the $k^{th}$ GP. We will need a way of evaluating a location with respect to each parameter of the conditional posterior. Inspired by \cite{roy2008}, we propose looking at the $u$-quantile of the conditional posterior $\pi(\alpha | \bm\gamma, \bm y) = $ $\mathcal F(\alpha | \psi_1(\bm\gamma), \cdots \psi_r(\bm\gamma)$. For most choices of $u$, this is guaranteed to be a meaningful combination of the $r$ $\psi_j$ parameters which gives us a way to evaluate the improvement with respect to each GP simultaneously. Secondly, by targeting a particular quantile of the conditional posterior, we are in fact targeting a tail of the modularization posterior. 
\end{itemize}

In the univariate case, $A \subset \mathbb R$, we propose the acquisition function
\begin{align}
\label{eq:crit}
I(\bm\gamma|\mathcal D) &= \text{Var}\Big(\hat\zeta_u(\bm\gamma|\mathcal D)\Big)\\[1.1ex]
\label{eq:critf}
&= \text{Var}\Big(\mathcal Q\left(u\big\vert \hat\psi_1(\bm\gamma|\mathcal D), \psi_2(\bm\gamma|\mathcal D), \cdots \psi_r(\bm\gamma|\mathcal D)\right)\Big)
\end{align}
where $\hat\zeta_u(\bm\gamma)$ is the $u^{th}$ percentile of $\mathcal F(\alpha | \psi_1(\bm\gamma), \cdots \psi_1(\bm\gamma))$ and $\mathcal Q$ is the quantile function corresponding to the distribution $\mathcal F$. The inclusion of $\hat{\cdot}$ symbols and $\mathcal D$ in the notation serves as a reminder that we are estimating these quantities using all available information $\mathcal D$. 

\subsection{Univariate Normal Distribution}
For instance, in the case where $\mathcal F$ corresponds to a univariate normal distribution, the $u$-quantile can be written as
\begin{equation}\label{eq:normquantile}
\zeta_u(\bm\gamma) = \mu(\bm\gamma) + \Phi^{-1}(u)\sigma(\bm\gamma),
\end{equation}
and the acquisition function in \cref{eq:crit} reduces to
\begin{equation}\label{eq:critnorm}
I_\text{norm}(\bm\gamma|\mathcal D) = \text{Var}\left(\hat\mu(\bm\gamma)|\mathcal D\right) + \left(\Phi^{-1}(u)\right)^2\text{Var}\left(\hat\sigma(\bm\gamma)|\mathcal D\right),
\end{equation}

There may be cases where a normal distribution assumption for $\mathcal F$ is inappropriate, and another parametric form is desired. For instance, if expert opinion dictates that the support of $\alpha$ is bounded on an interval then a Beta distribution may be specified for $\mathcal F$. Similarly, if the parameter of interest should be strictly positive, then perhaps a Weibull or Gamma distribution should be specified. We note that MC can always be used to approximate the acquisition function. If the quantile function $\mathcal Q$ can be written in closed form, then analytic calculations can be used to produce an approximation to \cref{eq:crit}. Two examples are given below.

\subsection{Weibull Distribution}

Suppose that
$$\mathcal F\left(\alpha \big| \lambda(\bm\gamma), \kappa(\bm\gamma)\right) = \frac{\alpha^{\kappa^{-1}(\bm\gamma)-1}}{\kappa(\bm\gamma)\lambda(\bm\gamma)^{\kappa^{-1}(\bm\gamma)}}\exp\left\{-\left(\alpha/\lambda(\bm\gamma)\right)^{\kappa^{-1}(\bm\gamma)}\right\}.$$
For notational compactness, we define $\hat\psi = \text{E}\left(\hat\psi(\bm\gamma|\mathcal D)\right)$ and $\tilde\psi = \text{Var}\left(\hat\psi(\bm\gamma|\mathcal D)\right)$ (for $\psi=\lambda,\kappa$).
Using the Delta method with a second order expansion we obtain
\begin{equation}
\label{eq:critweib}
\begin{aligned}
I_\text{Weibull}(\bm\gamma|\mathcal D) &= u_\star^{2\hat\kappa}\left\{\tilde\lambda\left(1+2\tilde\kappa\log^2(u_\star)\right)+\hat\lambda^2\tilde\kappa\log^2(u_\star)\left(1-\frac{\tilde\kappa}{4}\log^2(u_\star)\right)\right\} \\[1.2ex]
\text{with } &u_\star = -\log(1-u),
\end{aligned}
\end{equation}

\subsection{Multivariate Normal Distribution}

Returning to the multivariate case ($\bm\alpha \in A \subset \mathbb R^p$), we note that \cref{eq:crit} is no longer well defined. Since there are multiple $\alpha_i$ parameters, there is no clear definition for $\zeta_u(\bm\gamma)$. Instead, we consider a linear combination of the components
\begin{equation}
\label{eq:lincombo}
\alpha_0 = \sum_{i=1}^pt_i\alpha_i.
\end{equation}
The $t_i$ coefficients are constants, which can be fixed at one for simplicity or taken to be the square root of the marginal prior precision for $\alpha_i$ so that each term is weighted equally according to the prior. Now the $u^{th}$ percentile of $\alpha_0$ is a meaningful function of every parameter in $\mathcal F$,
\begin{equation}
\label{eq:quant}
\zeta_u(\bm\gamma) = \sum_{i=1}^pt_i\mu_i(\bm\gamma) + \Phi^{-1}(u)\left\{\sum_{i=1}^pt_i^2\sigma_i(\bm\gamma)^2 + 2\sum_{i=1}^{n-1}\sum_{j=i+1}^nt_it_j\sigma_{ij}(\bm\gamma) \right\}^{1/2}.
\end{equation}
The acquisition function of \cref{eq:crit} can now be used for this quantity. Due to the square root, $\zeta_u(\bm\gamma)$ is no longer a linear combination of independent GPs, and therefore finding $\text{Var}(\hat\zeta_u(\bm\gamma|\mathcal D))$ is no longer straightforward. Approximation of this variance via Monte Carlo integration is still a valid option but will lead to slower evaluation of $I(\bm\gamma)$ (especially if $p$ is large). Alternatively, we can apply the Delta method with a second order expansion. As before, we define $\hat\psi = \text{E}\left(\psi(\bm\gamma|\mathcal D)\right)$ and $\tilde\psi = \text{Var}\left(\psi(\bm\gamma|\mathcal D)\right)$.
\begin{equation}
\label{crit4}
\begin{aligned}
I_p(\bm\gamma \mid \mathcal{D}) &= \Bigg[ E(\delta_1^2) + \Phi^{-1}(u)^2 E(\delta_2) 
+ 2\Phi^{-1}(u)E(\delta_1)E\left(\delta_2^{1/2}\right)\Bigg] \\
&\quad - \Bigg[E(\delta_1) + \Phi^{-1}(u)E\left(\delta_2^{1/2}\right)\Bigg]^2, \\[1.2ex]
E(\delta_1) &= \sum_{i=1}^p t_i \hat\mu_i, \quad E(\delta_1^2) = E(\delta_1)^2 + \sum_{i=1}^p t_i^2 \tilde\mu_i, \\
E(\delta_2) &= \sum_{i=1}^p t_i^2 \hat\sigma_i^2 + 2 \sum_{i=1}^{p-1} \sum_{j=i+1}^p t_i t_j \hat\sigma_{ij}, \\[1.2ex]
E\left(\delta_2^{1/2}\right) &\approx E(\delta_2)^{1/2} - \frac{1}{2} 
\left(\frac{1}{16E(\delta_2)}\right)^{3/2} \Bigg(\sum_{i=1}^p t_i^4 \tilde{\sigma}_i^2 
+ 2 \sum_{i=1}^{p-1} \sum_{j=i+1}^p t_i t_j \tilde\sigma_{ij}\Bigg).
\end{aligned}
\end{equation}

Despite its appearance, \Cref{crit4} can be evaluated efficiently. 

\bibliography{biblio}